\documentclass[final]{siamltex}
\usepackage{epsfig,amsmath,amssymb}

\title{Asymptotic expansion for the normal implied volatility in local volatility models}

\author{Viorel Costeanu and Dan Pirjol\thanks{J.~P.~Morgan, New York, NY 10172 ({\tt pirjol@mac.com}).}}

\begin{document}

\maketitle

\begin{abstract}
We study the dynamics of the normal implied volatility in a local 
volatility model, using a small-time expansion in powers of maturity $T$. 
At leading order in this expansion, the asymptotics of the normal implied
volatility is similar, up to a different definition of the
moneyness, to that of the log-normal volatility. This relation 
is preserved also to order $O(T)$ in the small-time expansion, and differences
with the log-normal case appear first at $O(T^2)$.
The results are illustrated on a few examples of local volatility models
with analytical local volatility,
finding generally good agreement with exact or numerical solutions.
We point out that the asymptotic expansion can fail if applied naively
for models with nonanalytical local volatility, for example which have 
discontinuous derivatives. Using perturbation
theory methods, we show that the ATM normal implied volatility for such a model
contains a term $\sim \sqrt{T}$, with a coefficient which is
proportional with the jump of the derivative. 
\end{abstract}



\pagestyle{myheadings}
\thispagestyle{plain}
\markboth{V.~Costeanu and D.~Pirjol}{Normal implied volatility dynamics}

\section{Introduction}

Local volatility models have been introduced some time ago \cite{Dupire,DermanKani}, see \cite{JimG} for an
overview, in
order to model the price evolution of a financial asset in a way consistent with the 
known European option prices on that asset. 

According to the Gy\"ongy theorem \cite{Gyongy}, there is a
unique one-dimensional stock process which can reproduce a given terminal distribution
at each time horizon. This process is usually written in a way reminiscent of a
log-normal process
\begin{eqnarray}\label{sigD}
dS(t) = S(t) \sigma_D(S(t), t) dW(t) + S(t) \mu(t) dt
\end{eqnarray}
where $\sigma_D(S(t),t)$ is the so-called local volatility. In the most general
case, it depends both on the stock price $S(t)$, and explicitly on time $t$.
The process (\ref{sigD}) reduces to a simple log-normal evolution for the case
of a constant local volatility. 

The local volatility and drift can be determined 
empirically from the observed market prices of options~\cite{Dupire,DermanKani} using
the Dupire equation
\begin{eqnarray}\label{Dupire}
\partial_T C(K,T) = \frac12\sigma_D^2(K,T)  \partial_K^2 C(K,T)  
+ \mu(T)(C(K,T)-  \partial_K C(K,T)) 
\end{eqnarray}
The drift $\mu(T)$ is found from  the time dependence of the forward price $F(T)$ as
\begin{eqnarray}\label{lndrift}
\mu(T ) =  -\partial_T \log F(T)\,.
\end{eqnarray}
In principle this relation allows the determination of the local volatility in 
terms of option prices with all strikes and maturities. However, in practice this leads to 
numerical instabilities, and an equivalent relation is used, which replaces the option 
prices with the log-normal volatility \cite{BBF}. 

Many particular cases of local volatility $\sigma_D(S(t),t)$ have been considered
in the literature, such as for example the CEV model~\cite{CEV}, and
the quadratic volatility model~\cite{quad}.

The log-normal assumption implicit in (\ref{sigD}), and the quotation of the
implied volatility as log-normal volatility, are justified for equities, for which a negative asset value $S(t)<0$ is unphysical.
For interest rates this is not a restriction anymore, and in fact normal volatilities are commonly used in quoting
market volatilities as yield volatilities. There are a few advantages to using a normal volatility language for rates, the
most important of which is the fact that normal volatilities are more stable under shifts of 
the rates. Under certain conditions, calibrating a model 
to normal volatilities improves the calibration stability in time.

In this paper we consider the local volatility model for situations 
where a normal process is more natural, and replace the evolution equation 
(\ref{sigD}) with 
\begin{eqnarray}\label{sigDN}
dS(t) = \sigma_D(S(t), t) dW(t) + \mu(t) dt
\end{eqnarray}
In Section 2 we derive the Dupire equation for this process, and convert it to an expression for the
local volatility in terms of normal implied volatilities. In Section 3 we use asymptotic methods to derive an
expansion in powers of time for the
normal implied volatility. The leading term in this expansion is shown to have a very similar form to that
well-known  in the context of log-normal rates (the so-called
Berestecky-Busca-Florent (BBF) formula \cite{BBF}). The same result holds for the
term linear in time, and differences with the log-normal case appear first at quadratic
order in time.
The general
method is illustrated in Section 4 on a few examples of local volatility models. 
In Section 5 we consider the case of non-analytic local volatility functions,
and show on an explicit example that the asymptotic expansion can fail in such
cases. Using perturbation theory methods we prove that for local volatility 
models where the local volatility has a discontinuous derivative, the
normal implied volatility contains a nonanalytic term
$\sim \sqrt{T}$ which is proportional to the jump of the derivative.
This signals a generic failure of the asymptotic expansion for such
models, and underlines the need for care in their application.

\section{Dupire equation for the normal volatility}
\label{sec:2}

Consider the process $S_t$ driven by the local volatility model with local volatility $\sigma_D(S_t, t)$
\begin{eqnarray}\label{1}
dS_t = \sigma_D(S_t, t) dW_t +  \mu_t dt
\end{eqnarray}
We work in the risk-neutral measure, where the drift $\mu_t = r_t - D_t$ is the difference between the interest rate $r_t$, assumed to be
deterministic, and the dividend rate $D_t$, which is assumed to be paid continuously. 
Note the form of the drift term, which is different from the one usually adopted in local volatility models with a log-normal-type evolution
(\ref{sigD}).
This change is required by later convenience, and it will allow us to express $\sigma_D$ in terms
of the implied normal volatility.

This different choice affects also the relation between the drift $\mu_t$ and the forward stock price $F_T$, which reads 
\begin{eqnarray}
\mu(T ) =  \partial_T  F(T)\,.
\end{eqnarray}
Note the difference with the log-normal case Eq.~(\ref{lndrift}).

We will consider European options (calls and puts) on the asset $S_t$. The undiscounted price of a call with strike $K$ and maturity $T$
is given by the usual formula
\begin{eqnarray}
C(K,T) = \mathbb{E}[(S_T - K)_+]
\end{eqnarray}
The option prices can be quoted either in terms of log-normal and normal implied volatilities. We will consider here the normal implied 
volatility $\sigma_N(K,T)$, which is defined through the price of a call option assuming normally distributed
terminal stock price $S_T = F_T + \sigma_N \sqrt{T} X$, with $X\sim N(0,1)$ a normally distributed random 
variable. The result for the call price is~\cite{Bachelier}
\begin{eqnarray}\label{Bachelier}
C_N(K, T) = (F_T - K) N\Big(\frac{F_T-K}{\sigma_N \sqrt{T}}\Big) + 
\frac{1}{\sqrt{2\pi}}e^{-(K-F_T)^2/(2\sigma_N^2 T)} \sigma_N\sqrt{T}
\end{eqnarray}

We would like to find the forward equation satisfied by the prices of the call options $C(K,T)$ under the process (\ref{1}). This will
give the generalization of the well-known Dupire equation to this case. We start by quoting the result
\begin{eqnarray}\label{DupireN}
\partial_T C(K,T) = \frac12\sigma_D^2(K,T)  \partial_K^2 C(K,T)  - \mu(T)  \partial_K C(K,T) 
\end{eqnarray}
which has a very similar form to the usual Dupire equation with the obvious substitution $K^2 \sigma_D^2(K,T) \to \sigma_D^2(K,T)$.
In addition, the form of the drift term is different, due to the different choice for the drift term in (\ref{1}).

{\em Proof.} The proof of (\ref{DupireN}) proceeds in close analogy with the proof of the usual Dupire equation (\ref{Dupire}), see e.g. \cite{JimG}.
Start with the Fokker-Planck (FP) equation for the pdf of $S(T)$ starting with the initial condition $\varphi(S,0) = \delta(S)$
\begin{eqnarray}\label{FP}
\partial_T \varphi(S,T) = \frac12 \partial_S^2 [ \sigma_D^2(S) \varphi(S,T) ] - \partial_S [\mu(t) \varphi(S,T) ]
\end{eqnarray}

The call price is expressed in terms of the pdf $\varphi(S,T)$ as
\begin{eqnarray}
C(K,T) = \int_0^\infty dS(S - K)_+ \varphi(S,T)
\end{eqnarray}
Taking a derivative with respect to time and using the FP equation (\ref{FP}) we find
\begin{eqnarray}\label{sigmaD}
\partial_T C(K,T) &=& \int_K^\infty dx (x-K) \partial_T \varphi(x,T) \\
&=& \int_K^\infty dx (x-K) \{ -\partial_x (\mu(T) \varphi(x,T)) + \frac12 \partial_x^2 ( \sigma_D^2 \varphi(x,T)) \}\nonumber
\end{eqnarray}

The first integral is equal to 
\begin{eqnarray}
\int_K^\infty dx (x-K) \partial_x (\mu(T) \varphi(x,T)) = - \mu(T) \int_K^\infty 
dx  \varphi(x,T)
\end{eqnarray}
and the second one is
\begin{eqnarray}
\int_K^\infty dx (x-K) \{ - \partial_x^2 ( \sigma_D^2 \varphi(x,T)) =
\sigma_D^2(K,T) \varphi(K,T)
\end{eqnarray}
where we integrated by parts a few times, and assumed that the pdf $\varphi(x,T)$ vanishes
sufficiently fast as $x\to \infty$.

Putting everything together we obtain for (\ref{sigmaD})
\begin{eqnarray}
\partial_T C(K,T) &=&
\frac12 \sigma_D^2(K,T) \varphi(K,T) + \mu(T) \int_K^\infty dx \varphi(x,T) \nonumber \\
&=& \frac12  \sigma_D^2(K,T) \partial_K^2 C(K,T) - \mu(T) \partial_K C(K,T) \nonumber
\end{eqnarray}
where in the second line we used $\partial_K C(K,T) = - \int_K^\infty dx \varphi(x,T)$.
This completes the derivation of the Dupire equation (\ref{DupireN}) for the process (\ref{1}).

The relation (\ref{DupireN}) allows the determination of the local volatility $\sigma_D(S,T)$ from the market prices of the
options with different strikes and maturities. As mentioned, this determination is not very stable numerically,
due to the appearance of $0/0$ instabilities for short times and near ATM strikes. For this reason an alternative
approach is preferable, which replaces the option price with the respective implied volatililty.

In the following we will derive such a relation in terms of the implied normal volatility. We follow closely the derivation
in \cite{JimG}, adapting it to the problem at hand. 

We start by expressing the Bachelier option price in terms of the new independent variables 
\begin{eqnarray}
y = K- F_T\,,\qquad w = \sigma_N^2 T
\end{eqnarray}
We have
\begin{eqnarray}\label{Cyw}
C_N(y,w) = -y N(-\frac{y}{\sqrt{w}}) + \frac{1}{\sqrt{2\pi}}e^{-y^2/(2w)} \sqrt{w}\,.
\end{eqnarray}
Setting the call price $C(K,T)$ equal to the Bachelier price imposes a special functional
dependence of $w$ on $y$ and time
\begin{eqnarray}
C(K,T) = C(y+F_T,T) = C_N\Big(y, w(y,T)=[\sigma_N^2(y+F_T, T)T]\Big)
\end{eqnarray}

The derivatives with respect to $K,T$ in the Dupire formula can be replaced with derivatives with respect to $y,w$
of the Bachelier price $C_N(y,w)$, allowing for $y-$dependence in the $w$ variable due to the smile
\begin{eqnarray}
&& \partial_T C(K,T) = (\partial_T w)_K (\partial_w C(y,w))_y  + (\partial_T y)_K (\partial_y C(y,w))_w \\
&&\hspace{2cm}  =  \Big(\frac{dw}{dT}\Big)_K \partial_w C_N(y,w) - \mu_T \partial_y C_N(y,w) \nonumber\\
&& \partial_K C(K,T) = \partial_y C(y,w) = \partial_y C_N(y,w) + \frac{dw}{dy} \partial_w C_N(y,w) \\
&& \partial_K^2 C(K,T) = \partial_y^2 C_N + 2 \frac{dw}{dy}\partial^2_{yw}C_N + \frac{d^2w}{dy^2} \partial_w C_N + (\frac{dw}{dy})^2 \partial^2_w C_N 
\end{eqnarray}
The higher derivatives of $C_N(y,w)$ can be simplified with the help of the relations
\begin{eqnarray}
&& \partial_y^2 C_N = 2 \partial_w C_N \\
&& \partial_{yw}^2 C_N = -\frac{y}{w} \partial_w C_N \\
&& \partial_w^2 C_N = (\frac{y^2}{2w^2} - \frac{1}{2w}) \partial_w C_N 
\end{eqnarray}

Using these identities, the Dupire equation gives the following expression for the local volatility as a function of the normal volatility
\begin{eqnarray}
\sigma_D^2(K=y+F_T,T) = \frac{(dw/dT)_{K=y+F_T} + \mu_T (dw/dy)}{1 - \frac{y}{w}(\partial_y w) + \frac12 (\partial_y^2 w) + (\frac{y^2}{4w^2} - \frac{1}{4w}) (\partial_y w)^2}
\end{eqnarray}
The numerator of this expression can be written even simpler as the total time
derivative of $w(y,T)$: $\frac{d}{dT}w(y+F_T,T) = (dw/dT)_{K=y+F_T} + \mu_T (dw/dy)$. 
This result extends the well-known formula for the log-normal case (see e.g. Eq.(1.10) in \cite{JimG})  to the case of the normal implied volatility. 
In the log-normal case an additional term is present in the denominator, equal to $-\frac{1}{16}(\partial_y w)^2$~\cite{JimG}.
Converting from $w$ to the normal implied volatility $\sigma_N(y,T)$ we find the more 
explicit result 
\begin{eqnarray}\label{sigDN1}
\sigma_D^2(K,T) &=& \frac{\sigma_N^2(y,T) + T \partial_T \sigma_N^2(y,T) + \mu_T T \partial_y \sigma_N^2(y,T)}{1 - 2\frac{y}{\sigma_N(y)}\partial_y \sigma_N(y) + \frac{y^2}{\sigma_N(y)^2} (\partial_y \sigma_N(y))^2 + \frac12 T \partial_y^2 \sigma_N^2(y) - \frac{T}{4\sigma_N^2}(\partial_y\sigma_N^2(y))^2] }\nonumber\\
&=& \frac{\sigma_N^2(y,T) + T \partial_T \sigma_N^2(y,T) + \mu_T T (\partial_y \sigma_N^2(y,T))}
{(1 - \frac{y}{\sigma_N(y)}\partial_y \sigma_N(y))^2  +  T \sigma_N(y) \partial_y^2 \sigma_N(y) ] }
\end{eqnarray}

It is instructive to compare these results with the corresponding results for 
the log-normal case. The corresponding expression reads (note that the natural 
variable for this case is the log-strike $x = \log(K/F_T)$, instead of $y=K-F_T$)
\begin{eqnarray}\label{sigDNlogn}
\sigma_D^2(K=F_T e^x,T) &=&
 \frac{\sigma_{BS}^2(x,T) + T \partial_T \sigma_{BS}^2(x,T) + \mu_T T (\partial_x \sigma_{BS}^2(x,T))}
{(1 - \frac{x}{\sigma_{BS}(x)}\partial_x \sigma_{BS}(x))^2  +  T \sigma_{BS}(x) \partial_x^2 \sigma_{BS}(x) - \frac14 \sigma_{BS}^2 T^2 (\partial_x \sigma_{BS}(x))^2}\nonumber \\
\end{eqnarray}
We observe that the $O(T)$ terms are the same, but the $O(T^2)$ term present here is missing in the 
normal case Eq.~(\ref{sigDN1}). 
Apart from this, the two expressions are identical, up to the different definitions of the independent variables $x$ and $y$, as pointed out above.

We will neglect the possibility of explicit time dependence of the local volatility, and consider
only the case of a time-homogeneous local volatility, which depends on $S_t$ alone, such as in the CEV models and the
shifted log-normal model to be discussed below. 

In the short time limit $T\to 0$ the equation (\ref{sigDN1}) simplifies by dropping the terms
proportional to $T$, and setting $F_T \to F_0$ on the left-hand side. The equation can be solved
in closed form and the asymptotic normal implied volatility is given by
\begin{eqnarray}\label{BBFN}
\sigma_{N,0}(y) = \frac{y}{\int_0^y  \frac{dz}{\sigma_D(z)}} =  \frac{K-F_0}{\int_{F_0}^K  \frac{dL}{\sigma_D(L)}} \,. 
\end{eqnarray}
This result is the analog of a corresponding asymptotic expression obtained in the log-normal case in
\cite{BBF}.
In this limit the equation (\ref{sigDNlogn}) can be solved with the well-known result
given by the BBF formula
\begin{eqnarray}\label{BBF}
\sigma_{BS,0}(x) = \frac{x}{\int_0^x  \frac{dz}{\sigma_D(z)}} =  \frac{\log(K/F_T)}{\int_{F_T}^K  \frac{dL}{L\sigma_D(L)}} \,. 
\end{eqnarray}
This expresses the fact that, in the asymptotic short-time limit, the log-normal implied 
volatility is the harmonic average of the local volatility.

For strikes close to the ATM point $|y| \ll F_0$, the integral in the denominator of (\ref{BBFN})
can be approximated by the
rectangular rule, which gives a very simple approximation for the normal implied volatility near the ATM point
\begin{eqnarray}
\sigma_N(K ) \sim \sigma_D(\frac12(K + F_0)) - \frac{1}{24}(K- F_0)^2 \sigma_D^2(\frac12(K+F_0)) 
\Big(\frac{1}{\sigma_D(S)}\Big)''|_{S=\tilde K}
\end{eqnarray}
with $\tilde K$ an undetermined point in the $(F_0,K)$ interval. The error estimate corresponds to the 
Newton-Cotes formula of degree 2~\cite{AS}.

This extends the familiar Hagan-Woodward approximation \cite{HW,SABR} for the implied volatility in the log-normal case to the
normal case, and allows the use of intuition familar from the former case also to the latter. For instance, the ``local volatility skew is twice
the implied volatility skew'' statement holds for both cases, with the obvious correspondence between the local volatilities in the
two cases. 

The asymptotic result for the normal implied volatility $\sigma_0(y)$ and its derivatives are
related to the local volatility $\sigma_D(y)$ at the point $y=K-F_0=0$ as
\begin{eqnarray}\label{sig0atzero}
\sigma_0(0) &=& \sigma_D(0) \,,\qquad \sigma'_0(0) = \frac12 \sigma'_D(0) \\
\sigma''_0(0) &=& \frac13 \sigma''_D(0) - \frac{(\sigma'_D(0))^2}{6\sigma_D(0)}\nonumber \\
\sigma'''_0(0) &=& \frac14 \sigma'''_D(0) -  \frac{\sigma'_D(0)\sigma''_D(0)}{2\sigma_D(0)}
+ \frac{(\sigma'_D(0))^3}{4\sigma_D^2(0)}\,.\nonumber
\end{eqnarray}
These relations can be easily obtained by taking the derivatives of the BBF formula
(\ref{BBFN}) at the ATM point $y=0$.

\section{Asymptotics of the normal volatility}
\label{sec:3}

In this section we consider the inverse problem of determining the normal implied volatility $\sigma_N(y,T)$ for a given local volatility model,
with a given local volatility $\sigma_D(y,T)$. This problem is equivalent to the that of solving the evolution equation (\ref{1}) for the local
volatility model.There are several approaches to this problem in the literature.

\begin{itemize}

\item The method of asymptotic expansion in time $T$ of the Dupire equation. This was already hinted at in the previous Section, where we
derived the leading term of this expansion (\ref{BBF}). This method can be extended to higher orders in $T$, see \cite{HLbook,Asympt}.

\item The heat kernel expansion~\cite{HLpreprint,HLbook,Asympt}. This method uses the expansion of the Green's function for the Fokker-Planck equation (\ref{FP}),
exploiting its similarity to the heat equation. This allows the use of asymptotic methods developed for the parabolic partial differential
equation with space-dependent coefficients \cite{Yoshida,Varadhan}. 


\end{itemize}

The simplest local volatility model corresponds to a time-homogeneous local volatility 
$\sigma_D(S(t))$, which depends only on $S(t)$, but not explicitly on time. 
We will restrict ourselves to this case in the following. 

As mentioned above, the time-dependence of the forward $F_T$ introduces time dependence
even in this case. The drift term in the evolution equation can be eliminated by defining the 
forward asset price for a maturity $T$
\begin{eqnarray}
F(t,T) = S(t) + \int_t^T ds \mu(s)
\end{eqnarray}
This follows the driftless process
\begin{eqnarray}
dF(t,T) = \sigma_D(F(t,T) - \int_t^T ds \mu(s), t) dW(t)
\end{eqnarray}
This shows that, even if the local volatility does not depend explicitly on time, the evolution of 
$F(t,T)$ has explicit time-dependence introduced through the drift term \cite{Asympt}.

In the following we will use as starting point the equation (\ref{sigDN}) and construct the solution as an 
expansion in powers of time $T$.
There are two ways to solve the equation (\ref{sigDN1}) for $\sigma_N(y,T)$: working
at fixed strike $K$, or at fixed $y=K-F_T$. In both cases the time dependence is made
explicit by expanding in powers in $T$. Ultimately, we would like to find the
expansion of the implied volatility as a function of strike 
\begin{eqnarray}
\sigma_N(K,T) = \sigma_0(K) +  \sigma_1(K) T + \sigma_2(K) T^2 + \cdots
\end{eqnarray}
Working at fixed $y$ one would have to deal with additional
time dependence on the left-hand side of (\ref{sigDN1}) introduced through 
$F_T$ in $\sigma_D(K=y+F_T)$ which will
have to be expanded too. In addition, the form of the final result
expressed as function of strike would have the form $\sigma_N(K,T) = 
\sum_i \sigma_i(y+F_T) T^i$, and would mix orders in $T$.

Both these inconveniences are avoided by working at fixed $K$. We will thus treat Eq.~(\ref{sigDN1})
as a differential equation in $K$, at each order in $T$. This requires that we make explicit the time
dependence in the factor $y = K - F_T$ appearing in the denominator of (\ref{sigDN1}). The simplest 
way to do this is to redefine $y = K - F_0$, and replace $y \to y - (F_T - F_0) = 
y - \mu_0 T - \frac12 \mu_1 T^2 - \cdots$ in the denominator of  (\ref{sigDN1}). The resulting time
dependence is also made explicit by expanding in $T$.
Here we expanded also the drift $\mu_T$ in a power series in $T$
\begin{eqnarray}
\mu_T = \mu_0 + \mu_1 T + \mu_2 T^2 + \cdots
\end{eqnarray}
corresponding to the expansion of the forward 
\begin{eqnarray}
F_T = F_0 + \int_0^T dt \mu_t = F_0 + \mu_0 T + \frac12 \mu_1 T^2 + \cdots
\end{eqnarray}

The resulting expansion of the expression on the right-hand side of (\ref{sigDN1}) in powers of $T$ has a 
remarkably simple form to all orders in $T$
\begin{eqnarray}\label{expsigD}
\sigma_D^2(y) = \frac{\sigma_0^2(y)}{(1-\frac{y}{\sigma_0(y)}\sigma'_0(y))^2} \Big\{
1 + \sum_{j=1}^\infty T^j 
\Big[ \frac{2y}{1-\frac{y}{\sigma_0(y)}\sigma'_0(y)}  \Big(\frac{\sigma_j(y)}{\sigma_0(y)}\Big)' + 2(j+1) \frac{\sigma_j}{\sigma_0} + H_j(\sigma_{k<j}(y),y)\Big]\Big\}\nonumber\\
\end{eqnarray}
Recall that in this equation and below we use the definition $y = K - F_0$.
For simplicity, we denoted $\partial_y$  derivatives with respect to $y$ as primes. The function $H_j(y)$ depends only on $\sigma_0(y), \sigma_1(y), \cdots , \sigma_{j-1}(y)$, but not on $\sigma_j(y)$.
It depends also on the drift terms $\mu_k$ with $k<j$. 

The functional form (\ref{expsigD}) implies that the expansion terms $\sigma_j(y)$ can be 
determined recursively, starting with $\sigma_0(y)$ which is given by the BBF formula
\begin{eqnarray}
\sigma_{0}(y) &=& \frac{y}{\int_0^y  \frac{dz}{\sigma_{D}(z)}} =  \frac{K-F_0}{\int_{F_0}^K  \frac{dL}{\sigma_{D}(L)}} 
\end{eqnarray}

Assuming that all $\sigma_k(y)$ with $k<j$ are known (and thus the function $H_j(y)$ is known), 
the coefficient $\sigma_j(y)$ can be found by solving the differential equation 
obtained by equating the terms of $O(T^j)$ on the both sides of (\ref{expsigD})
\begin{eqnarray}\label{odesigmaj}
\frac{2y}{1-\frac{y}{\sigma_0(y)}\sigma'_0(y)}  \Big(\frac{\sigma_j(y)}{\sigma_0(y)}\Big)' + 
2(j+1) \frac{\sigma_j}{\sigma_0} + H_j(\sigma_{k<j}(y),y) = 0
\end{eqnarray}

The solution of the linear differential equation (\ref{odesigmaj}) can be 
found by the method of the variation of constants
\begin{eqnarray}
\sigma_j(y) = \sigma_0(y) \Big( \frac{\sigma_0(y)}{y}\Big)^{j+1} C_j(y)
\end{eqnarray}
where $C_j(y)$ is determined by the equation
\begin{eqnarray}
 \frac{2y}{1-\frac{y}{\sigma_0(y)}\sigma'_0(y)} \Big( \frac{\sigma_0(y)}{y}\Big)^{j+1} C'_j(y) + H_j(y) = 0
\end{eqnarray}
The solution is
\begin{eqnarray}\label{Cint}
C_j(y) &=& C_j(0) - \int_0^y \frac{dz}{2z} \Big(1 - \frac{z}{\sigma_0(z)} \sigma'_0(z)\Big)  
\Big( \frac{z}{\sigma_0(z)}\Big)^{j+1}  H_j(z)\\
&=& 
C_j(0) - \frac12 \int_0^y dz  \frac{z^j}{\sigma_D(z)\sigma_0^{j}(z)} H_j(z)\nonumber\,.
\end{eqnarray}
The integration constant $C_j(0)$ is determined by the condition that $\sigma_j(y)$ does not have a 
singularity at $y=0$, or equivalently $K=F_0$. This requires that $C_j(0) = 0$.

The final form of the solution is obtained by putting together the two factors, and is 
\begin{eqnarray}\label{sigmajsolution}
\sigma_j(y) = \sigma_0(y) \Big( \frac{\sigma_0(y)}{y}\Big)^{j+1} 
\Big[ - \int_0^y dz  \frac{z^j}{2\sigma_D(z)\sigma_0^{j}(z)} H_j(z) \Big]
\end{eqnarray}

The value of the $j-$th coefficient at $y=0$ depends only on the lower order coefficients, and does not 
require an integration. To see this, let's examine the contributions of the different terms in the square 
bracket in the expansion (\ref{expsigD})
at the point $y=0$. The first term vanishes, and the other two cancel among each other. 
This gives a relation for $\sigma_j(0)$ in terms of the lower order coefficients
\begin{eqnarray}\label{sigjATM}
\sigma_j(0) = -\sigma_0(0) \frac{H_j(0)}{2(j+1)}
\end{eqnarray}
Recall that due to the definition $y=K-F_0$ adopted in this section, the $y=0$ point corresponds to the
strike $K=F_0$, which coincides with the ATM point $K=F_T$ only if the drift vanishes.

This solves the recursion problem for $\sigma_j(y)$ in terms of the $\sigma_k(y)$ with $k<j$.
The only remaining problem is to find the function $H_j(y)$. This can be done by expanding the expression 
(\ref{sigDN}), and requires only algebraic manipulations.

\subsection{The solution for $\sigma_1(y)$}

Here we illustrate the general method outlined above on the example of the leading $O(T)$ correction 
to the BBF formula. The inhomogeneous term $H_1(y)$ is 
\begin{eqnarray}
H_1(y) &=& - \frac{\sigma_0(y)\sigma''_0(y)}{(1 - \frac{y}{\sigma_0(y)}\sigma'_0(y))^2} 
- 2\mu_0 \frac{\sigma'_0(y) \sigma_D(y)}{\sigma_0^2(y)}\Big( 1 - \frac{\sigma_0(y)}{\sigma_D(y)}\Big)
\end{eqnarray}

Substituting this into (\ref{Cint}) we get the integrals
\begin{eqnarray}
I_1 &=& \int_0^y dz \frac{z\sigma''_0(z)}{\sigma_0(z) - z \sigma'(z)} = - \int_0^y \frac{d[\sigma_0(z) - z\sigma'_0(z)]}{\sigma_0(z) - z \sigma'_0(z)} \\
&=& - \log( \sigma(y) - y \sigma'_0(y)) + \log(\sigma_0(0)) = 
- \log\Big(\frac{\sigma_0^2(y)}{\sigma_0(0)\sigma_D(y)}\Big)
\end{eqnarray}
and
\begin{eqnarray}
I_2 &=& \int_0^y dz \frac{z\sigma'_0(z)}{\sigma_0^3(z)}\Big( 1 - \frac{\sigma_0(y)}{\sigma_D(y)}\Big) = 
\int_{F_0}^K dz \Big( \frac{1}{\sigma_0(z)} - \frac{1}{\sigma_D(z)} \Big)^2
\end{eqnarray}

Combining everything gives for the first order (linear in time) correction to the BBF formula
\begin{eqnarray}\label{sig1N}
\sigma_{1}(y) &=& \frac{\sigma_{0}^3(y)}{y^2}\Big( -\frac12 \log \Big(\frac{\sigma_0^2(y)}{\sigma_D(y) \sigma_0(0)} \Big)
 + \mu_0 I_2 \Big) \\
&=& \frac{\sigma_{0}^3(y)}{y^2}\Big( -\frac12 \log \Big(\frac{\sigma_0^2(y)}{\sigma_D(y) \sigma_0(0)} \Big) +
\mu_0 
\int_{0}^y dz \Big( \frac{1}{\sigma_0(z)} - \frac{1}{\sigma_D(z)} \Big)^2 \Big) 
\nonumber
\end{eqnarray}
Expressed as a function of strike, the $O(T)$ correction to the normal implied volatility is
\begin{eqnarray}\label{sig1K}
\sigma_1(K) = \frac{\sigma_{0}^3(K)}{(K-F_0)^2}
\Big( -\frac12 \log \Big(\frac{\sigma_0^2(K)}{\sigma_D(K) \sigma_0(F_0)} \Big) +
\mu_0 
\int_{F_0}^K dz \Big( \frac{1}{\sigma_0(z)} - \frac{1}{\sigma_D(z)} \Big)^2 \Big) \nonumber\\
\end{eqnarray}
Of course, this correction is well-known and has been derived in \cite{HLbook} in the
context of the log-normal implied volatility, using a representation in terms of the process
for the forward stock price. As shown, at order $O(T)$ the asymptotic expansion of the
log-normal and normal implied volatilities are related by the simple replacement of the
log-strike variable $x = \log(K/S_0)$ with the variable $y = K -S_0$.
We have been unable to find an explicit result for the drift term
in the literature, apart from Ref.~\cite{Asympt}, see Eq.~(2.6) in this paper. Note however that
the second term in (\ref{sig1K}) is different from Eq.~(2.6) in \cite{Asympt} which has
$1/\sigma_0^2(z) - 1/\sigma_D^2(z)$ under the integral.

The $y=0$ value of the first subleading correction (\ref{sig1N}) can be obtained from (\ref{sigjATM}), and
depends only on the ATM local volatility and its derivatives
\begin{eqnarray}\label{sig1atzero}
\sigma_1(0) &=& \frac14 \sigma_0^2(0) \sigma''_0(0) + \frac12 \mu_0 \frac{\sigma'_0(0)}{\sigma_0(0)}(\sigma_D(0)-\sigma_0(0))\\
&=& \frac{1}{24} \sigma_D(0) [ 2\sigma_D(0) \sigma''_D(0) - (\sigma'_D(0))^2]\,.\nonumber
\end{eqnarray}
The ATM normal implied volatility up to $O(T)$ is given by
\begin{eqnarray}\label{sig1ATM}
\sigma_N^{(1)}(K=F_T) &=& \sigma_0(K=F_0+\mu_0 T) + \sigma_1(K=F_0 + \mu_0 T)T \\
&=&
\sigma_0(F_0) + (\sigma_1(F_0) + \mu_0 \sigma'_0(F_0))T + O(T^2) \,. \nonumber
\end{eqnarray}

The absence of a drift contribution to $\sigma_1(K=S_0,T)$ implies the following relation,
true for any local volatility model: the price of a call option with strike $K=S_0$ in the presence
of a constant drift $\mu$ is equal to the price of the same call option in the absence of the drift,
plus a known correction term $\frac12 \mu T$, up to terms quadratic in time
\begin{eqnarray}
C(K=S_0, T; \mu) = C(K=S_0, T;\mu = 0) + \frac12\mu T + O(T^2)
\end{eqnarray}

The higher derivatives of the $O(T)$ correction at the $K=S_0$ point are 
given by (up to terms proportional to the drift $\mu_0$)
\begin{eqnarray}\label{sig1p}
\sigma'_1(0) &=& \frac{1}{24}\Big( \sigma_D^2(0)\sigma'''_D(0) +
\sigma_D(0)\sigma'_D(0)\sigma''_D(0) - \frac12 (\sigma'_D(0))^3 \Big) \\
\label{sig1pp}
\sigma''_1(0) &=& \frac{1}{288}\Big( 12\sigma_D^2(0)\sigma''''_D(0) -
4\sigma_D(0)(\sigma''_D(0))^2 + 48 (\sigma'_D(0))^2\sigma''_D(0) - 13
\frac{(\sigma'_D(0))^4}{\sigma_D(0)}  \Big) \,.\nonumber\\
\end{eqnarray}

\subsection{The solution for $\sigma_2(y)$}

The second order coefficient can be computed using (\ref{sigmajsolution}) and is given by
\begin{eqnarray}
\sigma_2(y) = \sigma_0(y) \Big( \frac{\sigma_0(y)}{y}\Big)^{3} \Big[ - \int_0^y dz  \frac{z^2}{2\sigma_D(z)\sigma_0^{2}(z)} H_2(z) \Big]
\end{eqnarray}
where the inhomogeneous term in the equation for the $O(T^2)$ term is
\begin{eqnarray}
H_2(y) &=& 3\frac{\sigma_1^2(y)}{\sigma_0^2(y)} - 
\frac{4}{N^2} \frac{\sigma_1}{\sigma_0} \Big[-2Ny \Big(\frac{\sigma_1(y)}{\sigma_0(y)}\Big)' + \sigma_0(y) \sigma''_0(y) \Big] \\
&+& \frac{1}{N^4}\Big[-2Ny (\frac{\sigma_1(y)}{\sigma_0(y)})' + \sigma_0(y) \sigma''_0(y)\Big]^2 \nonumber \\
&-&   \frac{1}{N^2}\Big\{\Big[y\Big(\frac{\sigma_1(y)}{\sigma_0(y)}\Big)'\Big]^2 + 2Ny (\frac{\sigma_1(y)}{\sigma_0(y)})
\Big(\frac{\sigma_1(y)}{\sigma_0(y)}\Big)'+\sigma_1(y) \sigma''_0(y) + 
\sigma_0(y) \sigma''_1(y)\Big\}\nonumber\\
& & + H_2^{(\mu)}(y)\nonumber
\end{eqnarray}
where $N = 1 - \frac{y}{\sigma_0(y)} \partial_y \sigma_0(y) = \frac{\sigma_0(y)}{\sigma_D(y)}$.
The term $H_2^{(\mu)}(y)$ contains the dependence on drift, and is equal to
\begin{eqnarray}
H_2^{(\mu)}(y) &=& \mu_1 \frac{(\sigma_0^2(y))'}{\sigma_0^2(y)} + 
2\mu_0 \frac{(\sigma_0(y)\sigma_1(y))'}{\sigma_0^2(y)}
- \frac{4}{N}\mu_0^2 \Big(\frac{\sigma'_0}{\sigma_0}\Big)^2 \\
&-& 2\mu_0 \frac{\sigma'_0}{\sigma_0} \left[ \frac{\sigma_0\sigma''_0}{N^2} - \frac{2y}{N}
\Big(\frac{\sigma_1}{\sigma_0}\Big)' \right] + \frac{2\mu_0}{N^3} \frac{\sigma'_0(y)}{\sigma_0(y)}
- \mu_0^2 \frac{1}{N^2}\Big(\frac{\sigma'_0(y)}{\sigma_0(y)}\Big)^2
+ \mu_0\frac{2y}{N^2} \frac{\sigma'_0}{\sigma_0} \Big(\frac{\sigma_1}{\sigma_0}\Big)'\nonumber \\
&-&\mu_1 \frac{1}{N}\Big(\frac{\sigma'_0(y)}{\sigma_0(y)}\Big) - 
\mu_0\frac{2}{N}\Big(\frac{\sigma_1(y)}{\sigma_0(y)}\Big)'\,.\nonumber
\end{eqnarray}

The expression for $H_2(y)$ can be simplified using the equation satisfied by
$\sigma_1(y)$
\begin{eqnarray}
\frac{2y}{N}\Big(\frac{\sigma_1(y)}{\sigma_0(y)}\Big)' + 4\frac{\sigma_1(y)}{\sigma_0(y)}
+ H_1(y) = 0
\end{eqnarray}
We obtain
\begin{eqnarray}
H_2(y) = 3\Big(\frac{\sigma_1(y)}{\sigma_0(y)}\Big)^2 - 
\frac{(\sigma_0(y)\sigma''_0(y))^2}{4N^4} - \frac{\sigma_0(y)\sigma''_1(y)}{N^2}
+ H_2^{(\mu)}(y)
\end{eqnarray}
which gives the following result for the second order correction to the normal
implied volatility
\begin{eqnarray}
\sigma_2(y) &=& - \frac{\sigma_0^4(y)}{y^3}\int_0^y z^2 dz
\Big\{
\frac{3\sigma_1^2(z)}{2\sigma_D(z)\sigma_0^4(z)} -
\frac{\sigma_D^3(z)[\sigma''_0(z)]^2}{8\sigma_0^4(z)} -
\frac{\sigma_D(z) \sigma''_1(z)}{2\sigma_0^3(z)} + 
\frac{1}{2\sigma_D(z)\sigma_0^2(z)}H_2^{(\mu)}(z)
\Big\} \,.\nonumber \\
\end{eqnarray}
Its value at $K=S_0$ can be obtained from  Eq.~(\ref{sigjATM}) and does not require
an integration. Neglecting the drift terms, it is given by
\begin{eqnarray}
\sigma_2(K=S_0) = - \frac{\sigma_1^2(0)}{2\sigma_0(0)} + \frac{1}{24}\sigma_0(0)
[\sigma_0''(0)]^2 + \frac16 \sigma''_1(0)\,.
\end{eqnarray}
This can be expressed only in terms of the local volatility function
$\sigma_D(y)$ using Eqs.~(\ref{sig0atzero}), (\ref{sig1atzero}), (\ref{sig1pp}) for 
$\sigma_1(0)$ and $\sigma''_{0,1}(0)$.

\section{Examples}
\label{sec:4}

In this Section we apply the asymptotic method of Sec.~\ref{sec:3} to a few particular cases, 
in order to test the numerical convergence of the solution. We compare the asymptotic expansion
either to exact solutions, or to a numerical solution of the Dupire equation obtained by
solving it numerically in Mathematica \cite{Mathematica}.

\subsection{Shifted log-normal local volatility model}

Consider the local volatility model with shifted log-normal dynamics, and a constant drift term
\begin{eqnarray}\label{shiftlogn}
dS(t) = ( \sigma_0 + 2b S(t) ) dW(t) + \mu dt
\end{eqnarray}
This equation can be solved in closed form, and the solution reads
\begin{eqnarray}\label{Stpdf}
S(t) = (S_0 + \frac{\sigma_0}{2b} ) e^{2bW(t) - 2b^2 t} - \frac{\sigma_0}{2b}
+ \mu \int_0^t ds  e^{2b(W_t - W_s) - 2b^2 (t-s)}
\end{eqnarray}
The first term is log-normally distributed, while the second term, proportional
to the drift, has a more complicated distribution. Keeping only the first term,
the call price with strike $K$ and maturity $T$ can be expressed in terms of the
familiar Black-Scholes price $C_{BS}(K,F,\sigma_{BS},T)$ 
\begin{eqnarray}\label{shiftlnexact}
C(K,S_0,T) = C_{BS}(K=K + \frac{\sigma_0}{2b}, F=S_0+\frac{\sigma_0}{2b}, \sigma_{BS} = 2b, T)
\end{eqnarray}
The call price with nonzero drift has a more complicated expression, and is considered
in Appendix \ref{sec:appB}.

The leading asymptotic term for the normal implied volatility is obtained from (\ref{BBFN}) 
\begin{eqnarray}
\sigma_{0}(K) = \frac{K-F_T}{\int_{F_T}^K  \frac{dL}{\sigma_0 + 2bL}} = 
2b \frac{K-F_T}{\log\frac{\sigma_0 + 2bK}{\sigma_0 +2bS_0}}
\end{eqnarray}
The process (\ref{shiftlogn}) is invariant under the
simultaneous shifts $S\to S+\delta, \sigma_0 \to \sigma_0 - 2b\delta$, so
for simplicity we can choose $S_0=0$. The results for nonzero $S_0$ can be 
obtained by replacing $K\to K-S_0$.

Expanding in powers of $y=K-S_0$ one finds the leading order coefficient of the
asymptotic expansion
\begin{eqnarray}
\sigma_{0}(K) = \bar\sigma_0 + b y - \frac{b^2}{3\bar\sigma_0}y^2 + 
\frac{b^3}{3\bar\sigma_0^2}y^3 + O(y^4)
\end{eqnarray}
where $\bar\sigma_0=\sigma_0+2bS_0$.

\begin{figure}
    \centering
   \includegraphics[width=2.5in]{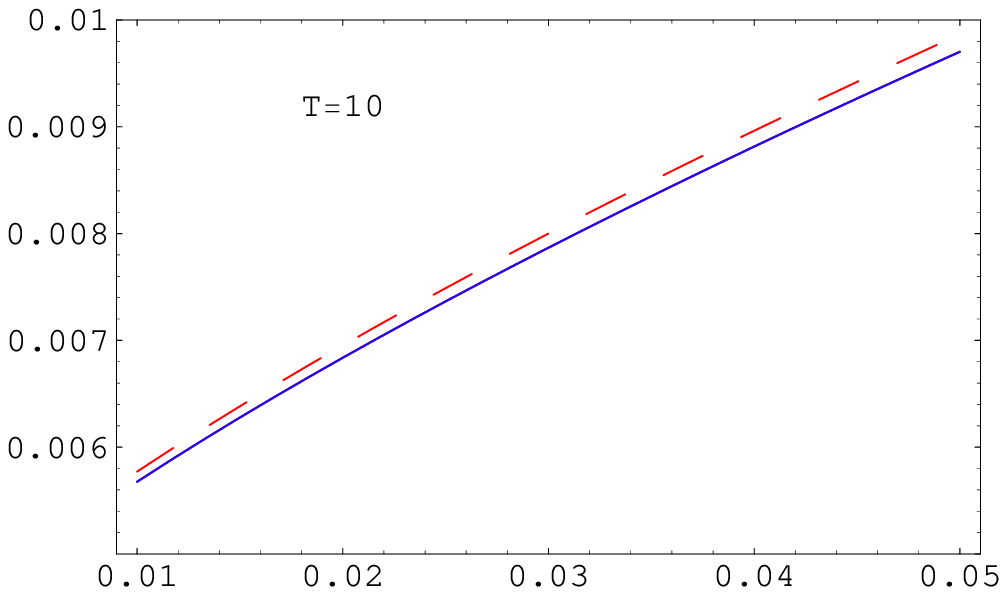}
   \includegraphics[width=2.5in]{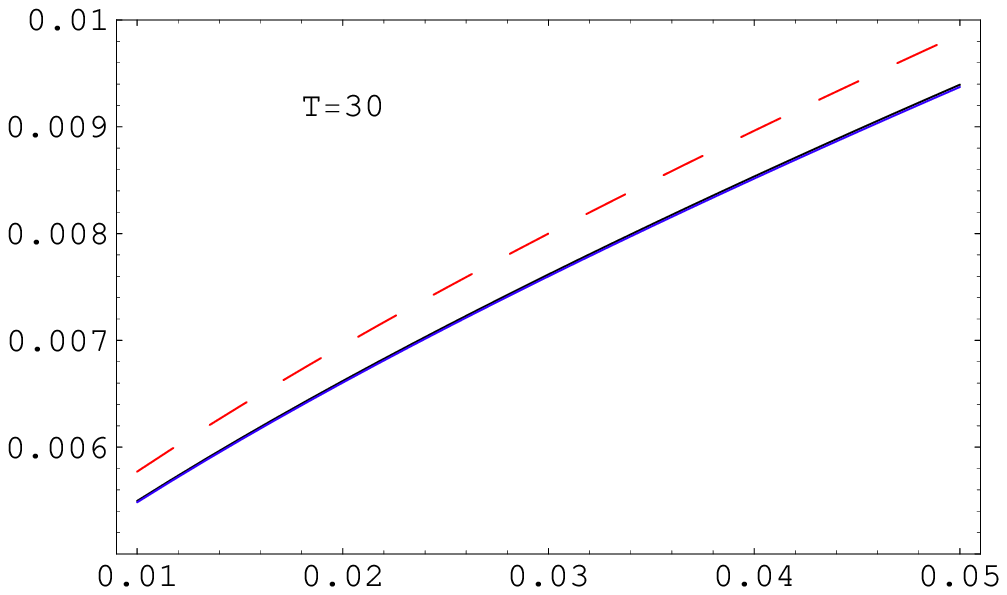}
    \caption{
The normal implied volatility $\sigma_N(K)-2bS_0$ for the shifted log-normal model
comparing the exact solution (black) with the asymptotic solution $\sigma_0(K)$
(red dashed), and the solution including $O(T)$ corrections (solid blue).
Model parameters: $S_0=3\%, \sigma_0 = 0.8\%, b=0.1$, and $T=10$ (left), $T=30$ (right).}
\label{Fig:1}
 \end{figure}

The correction of $O(T)$ is $\sigma_1(y)$, which can be obtained from (\ref{sig1N}). 
Its power expansion around the $y=0$ point is 
\begin{eqnarray}
\sigma_1(y) = - \frac16 b^2\bar \sigma_0 - \frac16 b^3 y + \frac{11b^4}{180\bar\sigma_0}y^2 + \cdots 
+ \mu_0(\frac{b^2}{3\bar\sigma_0} y - \frac{b^3}{3\bar\sigma_0^2} y^2 + O(y^3))
\end{eqnarray}

The ATM second order correction is obtained from (\ref{sigjATM}). Neglecting the drift, its expression is
\begin{eqnarray}
\sigma_2(0) &=& - \frac16 \sigma_0(0) H_2(0) = -\frac16 \sigma_0(y)
\Big[ 3 \frac{\sigma_1^2(y)}{\sigma_0^2(y)} - 5 \sigma_1(y) \sigma''_0(y) - 
\sigma_0(y) \sigma''_1(y) + (\sigma_0(y) \sigma''_0(y))^2 \Big] \nonumber \\
&=& -\frac16 \bar\sigma_0 \Big[
- \frac{3}{20} b^4
 \Big] = \frac{1}{40}\bar\sigma_0 b^4\,. 
\end{eqnarray}

Collecting all terms, the total ATM normal implied volatility in the (driftless) shifted 
lognormal model is given by an expansion in $b^2 T$, whose first three terms are 
\begin{eqnarray}
\sigma_N(K=S_0,T) = \bar\sigma_0 \Big(1 - \frac16 b^2 T + \frac{1}{40} b^4 T^2 + 
O((b^2T)^3)\Big)\,.
\end{eqnarray}

This can be compared with the exact solution (\ref{shiftlnexact}), which gives for the ATM normal volatility
\begin{eqnarray}
\frac{\sigma_N^{ATM}\sqrt{T}}{\sqrt{2\pi}} = \frac{\bar\sigma_0}{2b}[ N(b\sqrt{T}) - N(-b\sqrt{T})]
\end{eqnarray}
Expanding in powers of time we get
\begin{eqnarray}\label{sigmaN1B}
\sigma_N^{ATM}(T) = \bar\sigma_0 (1 - \frac16 b^2 T + \frac{1}{40}(b^2 T)^2   - \frac{1}{336} (b^2T)^3 + \frac{1}{3456} (b^2T)^4 + 
O(b^2T)^5 ) \nonumber\\
\end{eqnarray}
The first three terms agree with the coefficients from the asymptotic expansion obtained above.

To get a sense for the numerical accuracy of the expansion, we show in Figure \ref{Fig:1} the leading $O(T^0)$
term $\sigma_0(y)$ term in the asymptotic expansion (red dashed line), together with the first 
subleading correction $\sigma_0(y) + T \sigma_1(y)$ (blue curve), compared with the exact solution
(black curve). The agreement of the asymptotic expansion with the exact solution 
is very good, already at $O(T)$.

We show in Table 1 numerical results from the asymptotic expansion of the model, compared with the
exact solution. The model parameters have been chosen as $\sigma_0=3\%, b = 0.2$. 
The three columns show the error in the ATM normal implied volatility,
with respect to the exact result. At zeroth order in the $T$ expansion, the ATM implied normal vol is $\sigma_0 = 3\%$, but at large times it deviates from this value.
From Table~\ref{1Btable}, this deviation can be seen to be as large as $-0.5\%$ for $T=30$.

\begin{table}
\caption{\label{1Btable} 
Results for the ATM normal implied volatility in the shifted log-normal model, compared with the 
exact result, as a function of maturity $T$. The parameters of the model are 
$\bar\sigma_0 = 3\%, b = 0.2$.}
\begin{center}
\begin{tabular}{|c|c|ccc|}
\hline 
$T$ & $b^2 T$ & $\sigma_0-\sigma_{\rm ex}$ & $\sigma^{(1)}-\sigma_{\rm ex}$ & $\sigma^{(2)}-\sigma_{\rm ex}$ \\
\hline \hline
1 & 0.04 & $0.0199\%$ & $-0.0001\%$ & $0.0000\%$ \\
2 & 0.08 & $0.0395\%$ & $-0.0005\%$ & $0.0000\%$ \\
5 & 0.2 & $0.0971\%$ & $-0.0029\%$ & $0.0001\%$ \\
10 & 0.4 & $0.1885\%$ & $-0.0115\%$ & $0.0005\%$ \\
20 & 0.8 & $0.3562\%$ & $-0.0438\%$ & $0.0042\%$ \\
30 & 1.2 & $0.5058\%$ & $-0.0942\%$ & $0.0138\%$ \\
\hline \hline
\end{tabular}
\end{center}
\end{table}

Next we consider also the case with nonzero drift $\mu$. This can be included in the 
expressions for the coefficients of the asymptotic expansion as discussed in the previous
section, see Eq.~(\ref{sig1K}). In particular, the ATM normal implied volatility,
working to order $O(T^2)$, is given by (\ref{sig1ATM}), which for the case considered here, reads
\begin{eqnarray}\label{1Bsig1ATM}
\sigma_N(K=S_0 + \mu T) = \bar \sigma_0 + ( - \frac16 b^2 \bar \sigma_0 + b \mu) T + O(T^2)
\end{eqnarray}
where the $b\mu T$ term comes from the expansion of the leading order term
$\sigma_0(y)$ around $y = 0$.

We can check explicitly that this is reproduced by the exact solution of the model 
with constant drift, which is derived in the Appendix \ref{sec:appB}. In Eq.~(\ref{final})
we obtained the first two terms of the expansion of the call price with $K=S_0$
in powers of the drift $\mu$. 

The model (\ref{shiftlogn}) is related to the model in the Appendix \ref{sec:appB} by a 
simple mapping
\begin{eqnarray}
&& t = 4b^2 \tau \\
&& W_t = 2b \tilde W_\tau \\
&& x_t = S_\tau + \frac{\sigma_0}{2b} \\
&& \mu = \frac{\mu}{4b^2}
\end{eqnarray}
Then if $x_t$ satisfies the process $dx_t = x_t dW_t + \mu dt$, then $S_\tau$ will 
follow the process $dS_\tau = (\sigma_0 + 2bS_\tau) d\tilde W_\tau + \mu d\tau$.
 
In Appendix~\ref{sec:appB} it is shown that the price of an option with strike equal to 
the initial asset value $x_0$ is given by
\begin{eqnarray}
C(K=x_0,x_0,t) = x_0 \mbox{Erf}\Big(\frac{\sqrt{t}}{2\sqrt2}\Big) + \frac12\mu t + O(\mu^2)
\end{eqnarray}
The corresponding result for the model (\ref{shiftlogn}) is
\begin{eqnarray}\label{ATM1Bdrift}
C(K=S_0, S_0, T) = \frac{\sigma_0 + 2b S_0}{2b} 
\mbox{Erf}\Big(\frac{b\sqrt{T}}{\sqrt2}\Big) + \frac12 \mu T + O(\mu^2)\,.
\end{eqnarray}
 
From this result we can determine the normal implied volatility in the model (\ref{shiftlogn}) at the point $K=S_0$. This can be found by
comparing (\ref{ATM1Bdrift}) with the Bachelier formula with $\sigma_N(K=S_0,T) = \bar\sigma_0 (1 + c_1 T + O(T^2))$, which gives
$c_1 = - \frac16 b^2$. This is in agreement with the result Eq.~(\ref{1Bsig1ATM}) for $\sigma_1(K=S_0,T)$ 
obtained from the asymptotic expansion and confirms the absence of a term proportional to $\mu$ in 
$\sigma_1(K=S_0,T)$.

\subsection{Stochastic volatility inspired local volatility model}

As a second example, we derive the short-time asymptotics of the implied normal 
volatility in a local volatility model inspired by a stochastic volatility model. 
Consider the model
\begin{eqnarray}\label{SV}
dS(t) &=& \nu(t) \sigma_D(S(t)) dW_1(t) \\
d\nu(t) &=& \gamma \nu(t) dW_2(t)\,. \nonumber
\end{eqnarray}
where the two stochastic drivers have correlation $\mbox{corr }(W_1(t), W_2(t)) =\rho$. 
The initial condition is $S(0) = F_T, \nu(0) = 1$. 

We would like to use the methods of Sec.~\ref{sec:3} to find the short time asymptotics of the
normal volatility in this model. This can be done by relating first the model (\ref{SV})
to the SABR model \cite{SABR}, and then using the well-known asymptotic local volatility 
of the latter to find the equivalent local volatility of the model (\ref{SV}). 
We start by making the change of variable
\begin{eqnarray}\label{zdef}
z(S_t) = \int_{F_T}^{S_t} \frac{dy}{\sigma_D(y)}
\end{eqnarray}
which transforms the first equation (\ref{SV}) into
\begin{eqnarray}
dz = \nu(t) dW_1(t) + \mbox{ drift}
\end{eqnarray}

This is identical with the evolution of the log-price $x_t = \log(S_t/F_t)$ in the 
log-normal SABR model \cite{SABR}, so we can take over the asymptotic solution of the
SABR model for the short-time asymptotic limit \cite{SABR,SABR1}. 
The following one-dimensional 
process has the same terminal distribution of $x(t)$ as the two-dimensional stochastic 
volatility model at leading order in a short-time expansion \cite{SABR1,BBFstochvol}
\begin{eqnarray}
\sigma_{\rm eff}^2(z) = \gamma^2 z^2 - 2\rho \gamma z + 1 \,. 
\end{eqnarray}
Higher order terms in the asymptotic expansion of the SABR model have been obtained in 
\cite{HLpreprint,HLbook,SABR2,SABR3}.

The process for $z_t$ can be converted into a process for the original asset price 
$S(t)$ by an application of the Ito lemma
\begin{eqnarray}\label{SV1dim}
dS(t) = \sigma_D(S(t)) \sigma_{\rm eff}(z(S(t)) dW(t)
\end{eqnarray}
where the function $z(S)$ is given in (\ref{zdef}). 

We can use now the short-time asymptotics (\ref{BBFN}) applied to this one-dimensional model
to derive the leading asymptotics for the normal smile in the stochastic volatility model (\ref{SV})
\begin{eqnarray}\label{SVsol}
\sigma_{N,0}(K,T) &=& \frac{K-F_T}{\int_{F_T}^K \frac{dS}{\sigma_D(S) \sigma_{\rm eff}(z(S))}} =  \frac{K-F_T}{\int_{0}^{z(K)} \frac{dx}{\sigma_{\rm eff}(x)}} 
= \gamma  \frac{K-F_T}{D(\gamma z(K))} 
\end{eqnarray}
where
\begin{eqnarray}
D(x) = \log\frac{\sqrt{1-2\rho x +x^2}+ x - \rho}{1-\rho}
\end{eqnarray}
and 
\begin{eqnarray}
z(K) = \int_{F_T}^{K} \frac{dy}{\sigma_D(y)} \,.
\end{eqnarray}
The result (\ref{SVsol}) gives the short-time asymptotics for the normal implied volatility 
of the stochastic volatility model (\ref{SV}). 

We consider in the following the case of the normal SABR model with constant 
local volatility $\sigma_D(S) = \sigma_0$, and study the normal implied 
volatility of the local volatility model (\ref{SV1dim}). The argument of the 
local volatility depends on the variable $z(S(t),S(0)) = 1/\sigma_0(S(t)-S(0))$,
such that we obtain the local volatility model
\begin{eqnarray}\label{SABR1}
dS(t) = \sqrt{\sigma_0^2 - 2\rho \gamma\sigma_0 y + \gamma^2 y^2} dW(t)
\end{eqnarray}
with $y = S(t) - S(0)$.
Although the original
stochastic volatility model (\ref{SV}) is well-defined only for non-positive 
correlation $\rho \leq 0$ \cite{Jourdain}, we will use the local volatility 
(\ref{SV1dim}) for both positive and negative values of $\rho$, effectively 
ignoring its origin as a stochastic volatility model.

One can use now the results of Section~\ref{sec:2} to derive the normal implied volatility of this
model as an expansion in $T$. The leading order result is given in  Eq.~(\ref{SVsol}), and the
first subleading correction in Eq.~(\ref{sig1N}).

\begin{figure}
    \centering
   \includegraphics[width=2.5in]{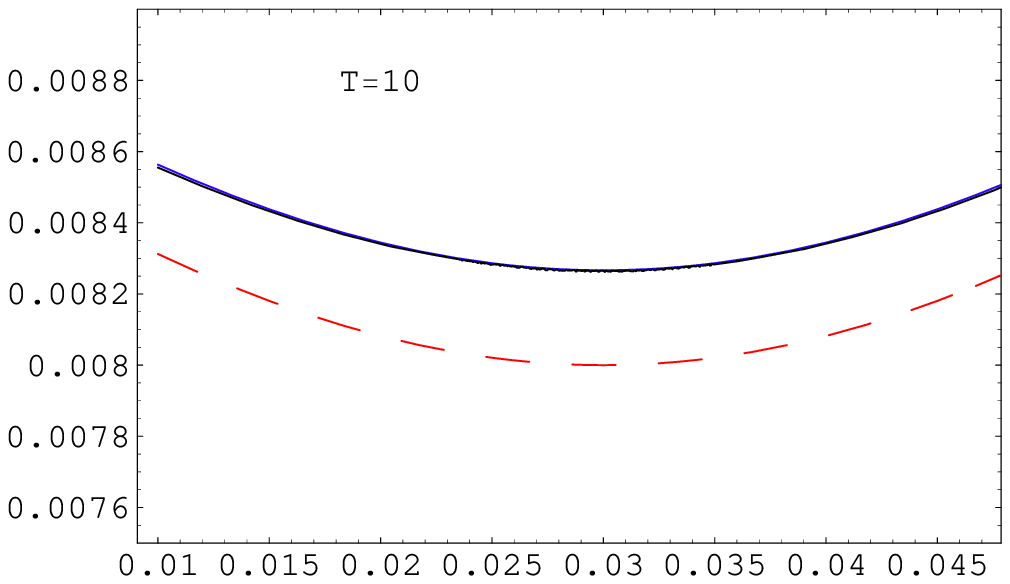}
   \includegraphics[width=2.5in]{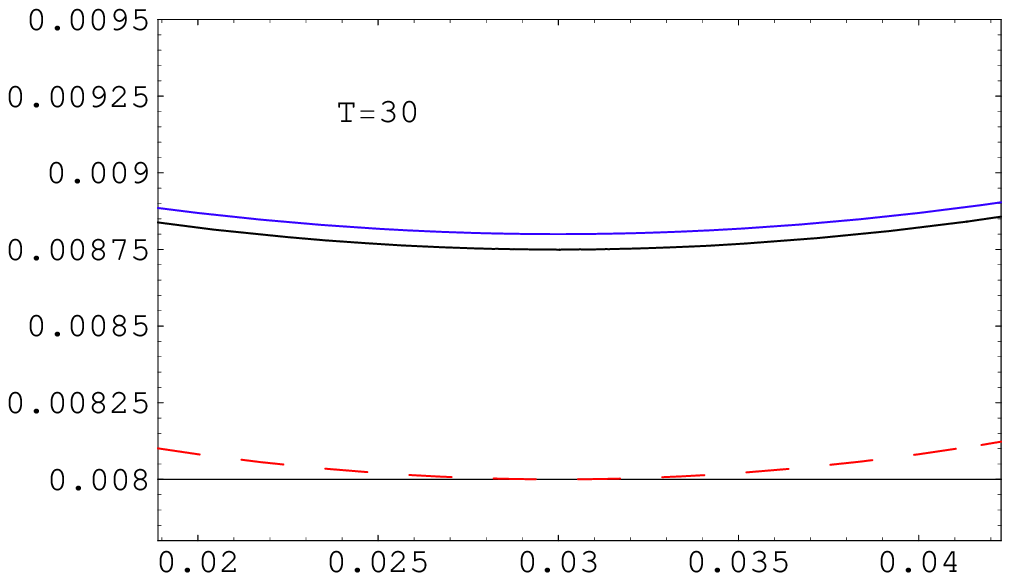}\\
   \includegraphics[width=2.5in]{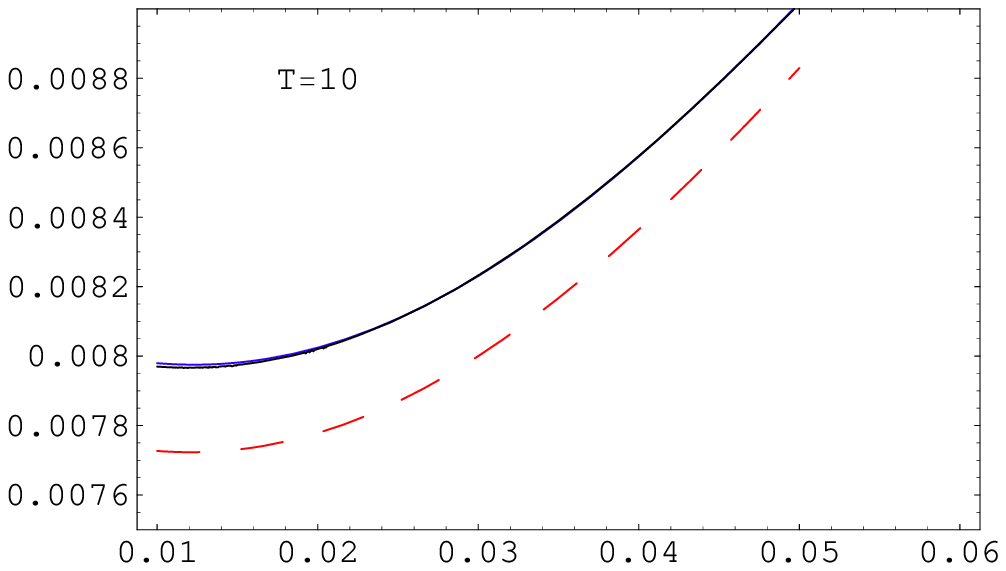}
   \includegraphics[width=2.5in]{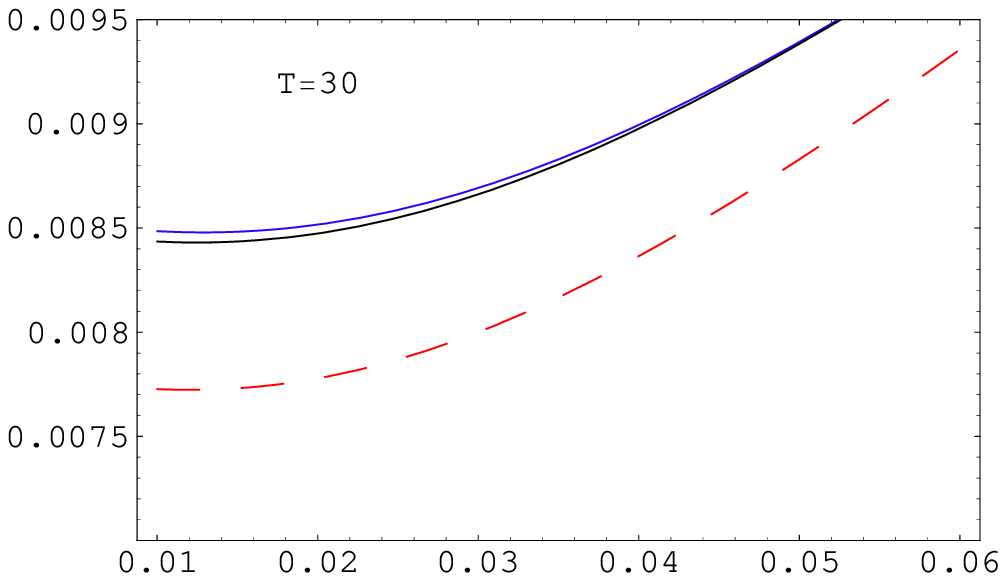}\\
   \includegraphics[width=2.5in]{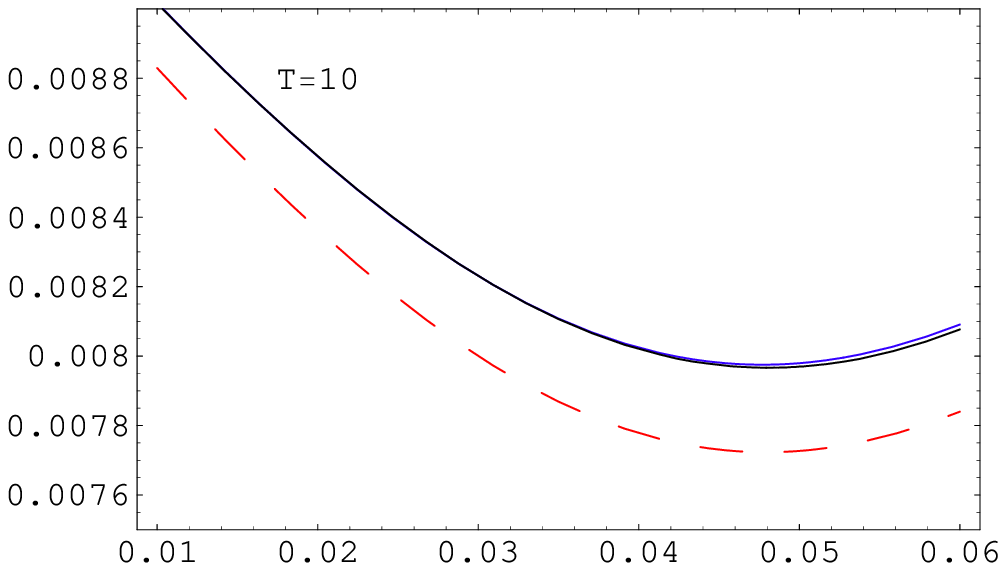}
   \includegraphics[width=2.5in]{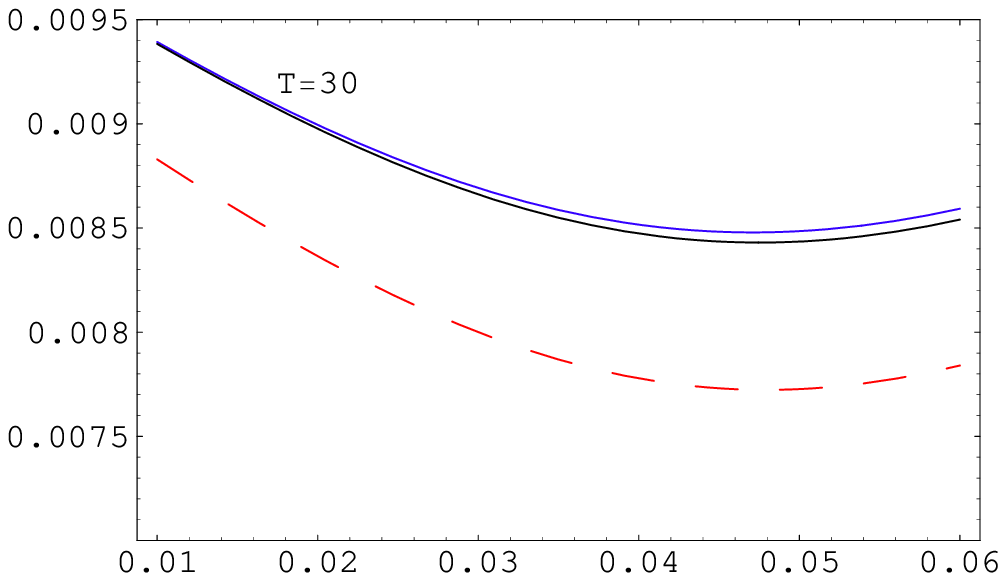}
    \caption{
The asymptotic expansion for the normal implied volatility of the SABR-like model (\ref{SABR1}) 
compared with the exact numerical 
solution (black curves) for two maturities: $T=10$ (left), and $T=30$ (right).
The red dashed curves show the
asymptotic solution $\sigma_0(K)$, and the blue curve shows the asymptotic
solution including the $O(T)$ correction.
Model parameters: $S_0=3\%, \sigma_0 = 0.8\%, \gamma=0.2$. The correlation is
$\rho = 0\%$ (upper plots), $\rho = 30\%$ (middle plots) and $\rho = -30\%$ (lower plots).  }
\label{Fig:2}
 \end{figure}

We show in Figure \ref{Fig:2} the results for the normal implied volatility in the
model (\ref{SABR1}) at leading order in $T$, and including the $O(T)$ correction 
$\sigma_1(y)$. 
These are compared with an exact numerical solution. We note that for moderate maturities the 
agreement obtained by keeping only the first subleading term is satisfactory.

\section{Nonanalytic local volatility}

The asymptotic expansion for the normal implied volatility in powers of $T$ 
presented in Sections 2 and 3 is applicable only for analytic local volatility
functions $\sigma_D(S)$. If $\sigma_D(S)$ is not analytic, the asymptotic
expansion fails, and new terms which are non-analytic in time can appear.

To illustrate this phenomenon, consider a model with local volatility which
has a discontinuous derivative
\begin{eqnarray}\label{2B}
\sigma_D(y) = \left\{
\begin{array}{cc}
\sigma_0 + 2b_L y\,, & y < 0 \\
\sigma_0 + 2b_R y\,, & y > 0 \\
\end{array}\right.
\end{eqnarray}. 

According to the
asymptotic expansion formula, the implied volatility $\sigma_0(y)$ (and all higher
order terms) depends only on the values
of the local volatility in the interval $(S_0, K)$, but not on its values outside
this interval. For this case, it implies that the implied volatility $\sigma_N(y)$
for $y>0(y<0)$ should
depend only on $b_R (b_L)$. We show in Fig.~3 the exact solution for the
implied volatility $\sigma_N(y)$, obtained using a numerical solution in
Mathematica (solid lines), and the leading asymptotic results $\sigma_0(K)$,
from the BBF formula (dashed lines). The three curves correspond to $b_L = -0.1$ (red),
$b_L = 0$ (blue), and $b_L=0.1$ (black), keeping $b_R=0.1$ fixed.
We observe that the implied volatility for $y>0$ depends strongly on the value
of $b_L$, which determines the local volatility outside the region $(S_0,K)$.
Although for $b_L = 0.1$ the exact result is in good agreement with the 
asymptotic expansion, the agreement becomes far worse for $b_L = 0, -0.1$.

In order to understand better the poor convergence of the asymptotic 
expansion in the case $b_L \neq b_R$, we consider in more detail the
particular case of the model (\ref{2B}) with
$- b_L = b_R \equiv b$, for which an exact solution is known~\cite{KS}. 
The exact solution 
for the ATM implied volatility can be written as a sum of two terms (see the Appendix
for a detailed proof). The first term
coincides with the solution of the shifted log-normal model considered above with $b = b_R$, 
and the second term gives the correction due to the discontinuity at $ y < 0$
\begin{eqnarray}\label{ATM2B}
\sigma(0,t) = \sigma_1(0,t) + \sigma_2(0,t)
\end{eqnarray}
with
\begin{eqnarray}
\sigma_1(0,t) &=& \sigma_0 \sqrt{\frac{\pi}{2}} \frac{1}{b\sqrt{t}} \mbox{Erf} 
(\frac{b\sqrt{t}}{\sqrt2} ) \\
\sigma_2(0,t) &=& \frac12 \sigma_0 e^{-\frac12 b^2 t} + \frac12 \sigma_0 \sqrt{\frac{\pi}{2}}
\Big[ b\sqrt{t} + b\sqrt{t} \mbox{Erf} 
(\frac{b\sqrt{t}}{\sqrt2} ) - \frac{1}{b\sqrt{t}}
\mbox{Erf} 
(\frac{b\sqrt{t}}{\sqrt2} ) \Big]
\end{eqnarray}

Expanding the exact implied volatility in powers of time, the first term
$\sigma_1(0,t)$ reproduces the result of the shifted log-normal model (\ref{sigmaN1B})
corresponding to the local volatility in the $y>0$ extrapolated to the entire
region $\sigma_D(y) = \sigma_0 + 2by$
\begin{eqnarray}
\sigma_1(0,t) &=& \sigma_0(1 - \frac16 b^2 t + \frac{1}{40} b^4 t^2 - \frac{1}{336} b^6 t^3 + \cdots) 
\end{eqnarray}
However, there is a second correction $\sigma_2(0,t)$, which has the time expansion
\begin{eqnarray}\label{sig22B}
\sigma_2(0,t) &=& \sigma_0 (\frac12 \sqrt{\frac{\pi}{2}} b\sqrt{t} + \frac13 b^2 t -
\frac{1}{30} b^4 t^2 + \frac{1}{280} b^6 t^3 + \cdots)
\end{eqnarray}
The term $\sigma_2(0,t)$ is not reproduced by the asymptotic expansion
discussed in Section 3. Even more
surprising, it contains a term of $O(\sqrt{t})$, which is not allowed by the
asymptotic expansion of the Dupire equation in volatility form. This term largely 
accounts for the large deviation from the asymptotic result in Figure 3.
The absence of the term $\sigma_2(0,t)$ and its nonanalytic dependence on time 
signal a failure of the asymptotic expansion for the model (\ref{2B}). 

\begin{figure}
    \centering
   \includegraphics[width=3in]{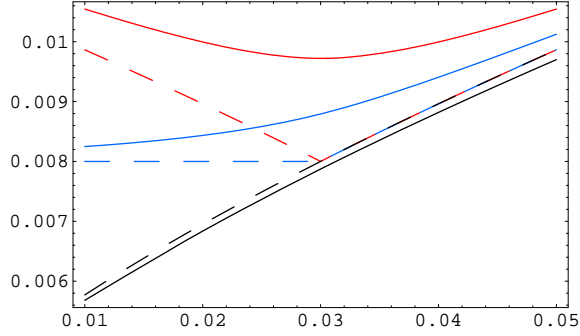}  
    \caption{
The normal implied volatility $\sigma_N(K)$ for the model (\ref{2B})
with parameters $S=3\%, \sigma_0 = 0.8\%, b_L =0.1 \mbox{(black}), 
0 \mbox{(blue)}, -0.1 \mbox{(red)}, b_R=0.1$ and $T=10$. 
The solid lines show exact numerical solutions of the Dupire equation, while the
dashed lines correspond to the asymptotic solution $\sigma_0(K)$.}
 \end{figure}

This failure appears to be generic for nonanalytic local volatility functions.
We prove next the following result: in any local volatility model with a local 
volatility function which has a jump in the derivative at the forward point
\begin{eqnarray}
\Delta (\sigma^2_D(F_T))' = \partial_K \sigma^2_D(K=F_T + \varepsilon) - \partial_K \sigma^2_D(K=F_T-\varepsilon)
\end{eqnarray}
the time expansion of the ATM normal implied volatility contains a term $\sim \sqrt{T}$ which
is nonanalytic in time, and which is proportional to the jump of the derivative
\begin{eqnarray}\label{sqrtTterm}
\sigma_N(F_T,T) = \sigma_D(F_T) + \frac{1}{16}\sqrt{\frac{\pi}{2}}\sigma_D(F_T)\sqrt{T}
\Delta(\sigma^2_D(F_T))' + O(T)
\end{eqnarray} 
This anomalous term reproduces the $O(\sqrt{T})$ term in Eq.~(\ref{sig22B}) for the 
model  (\ref{2B}), for which the jump of the derivative is $\Delta\sigma'_D(F_T)=4b$.

The proof of this result makes use of perturbation theory for linear 
operators~\cite{Kato,CCMN}.
We would like to solve the Dupire equation (\ref{DupireN}) which we repeat here for
convenience. This gives the price of a call option $C(K,T)$ at time $T$, and reads
\begin{eqnarray}
\partial_T C(K,T) = \frac12\sigma_D^2(K)  \partial_K^2 C(K,T)  
\end{eqnarray}
with the initial condition
$C(K,0) = \mbox{max}(K - S_0,0)$. For simplicity, we assumed a zero drift term $\mu_T=0$, 
and assumed the local volatility to be time-homogeneous.
This equation is identical with the heat equation with position dependent 
conductivity $\sigma_D(K)$, and the initial condition $C(K,0) = \mbox{max}(K - S_0,0)$.

The solution can be written formally  in operator form as
\begin{eqnarray}\label{solexp}
C(K,T) = \exp\Big(\hat L(K) T \Big) C(K,0) =
\exp\Big(\hat L_0 T + (\hat L-\hat L_0) T\Big) C(K,0)
\end{eqnarray}
where we denoted the differential operators
\begin{eqnarray}
\hat L_0(K) &=& \frac12\sigma_D^2(S_0) \partial_K^2 \\
\hat L(K) &=& \frac12\sigma_D^2(K) \partial_K^2 
\end{eqnarray}
The solution with $\hat L=\hat L_0$ is well-known, and can be written as a convolution
of the initial condition with the heat kernel
\begin{eqnarray}
G_0(x,t;y,0)  = \frac{1}{\sqrt{2\pi t}} \exp(-\frac{(x-y)^2}{2 t})
\end{eqnarray}
The solution of the unperturbed equation $\partial_T C_0(K,T) = \hat L_0(K) C(K,T)$ is
given by the convolution
\begin{eqnarray}
C_0(K,T) = \int_{-\infty}^\infty dy G_0(y,\sigma_0^2 T;y,0) \mbox{max}(y-S_0,0)
\end{eqnarray}
where we denoted for simplicity $\sigma_D(S_0) = \sigma_0$.
The integral can be computed in closed form, and  the result is the well-known Bachelier 
formula (\ref{Bachelier}).

We will treat the difference $\Delta \hat L \equiv \hat L-\hat L_0$ as a perturbation, 
and write the solution (\ref{solexp}) as an expansion in powers of $\Delta \hat L$
\begin{eqnarray}
C(K,T) = C_0(K,T) + C_1(K,T) + C_2(K,T) + \cdots
\end{eqnarray}
The term of zeroth order $C_0(K,T)$ is identical to the Bachelier result.
The term of first order in the perturbation is given by a triple integral
\begin{eqnarray}
C_1(K,T) = \int_{-\infty}^\infty dy \int_{-\infty}^\infty du 
\int_0^T d\tau
G_0(K,\sigma_0^2 T;u,\sigma_0^2 \tau) \Delta \hat L(u) G_0(u,\sigma_0^2 \tau; y,0) 
\mbox{max}(y-S_0,0)\nonumber \\
\end{eqnarray}

The action of the operator $\Delta \hat L(u)$ can be expressed as
\begin{eqnarray}
\Delta \hat L(u) G_0(u,\sigma_0^2 \tau; y,0) &=& 
\frac{\sigma_D^2(u) - \sigma_0^2}{\sigma_0^2}
\partial_\tau G_0(u,\sigma_0^2 \tau; y,0)\\
& & = \frac{\sigma_D^2(u) - \sigma_0^2}{\sigma_0^2}
\Big(-\frac{1}{2\tau} + \frac{(u-y)^2}{2\sigma_0^2 \tau^2}\Big) G_0(u,\sigma_0^2\tau; y,0)\,.\nonumber
\end{eqnarray}

Let us examine in some detail the dependence on time $T$ of the first order perturbation
$C_1(K,T)$ for the ATM case $K=S_0$. For simplicity, shift the origin of the $y$ axis such
that $S_0 = 0$. It is also useful to introduce rescaled variables 
\begin{eqnarray}
\bar y = \frac{y}{\sigma_0 \sqrt{T}}\,,\qquad
\bar u = \frac{u}{\sigma_0 \sqrt{T}}\,,\qquad
\lambda = \frac{\tau}{T}\,.
\end{eqnarray}
in terms of which the heat kernel is expressed as
\begin{eqnarray}
G_0(0,\sigma_0^2 T;u,\sigma_0^2 \tau) = \frac{1}{\sigma_0\sqrt{T}} G_0(0,1;\bar u,\lambda)
\end{eqnarray}

We have in terms of these variables
\begin{eqnarray}
C_1(0,T) &=& \sigma_0\sqrt{T}
\int_{-\infty}^\infty d\bar y  d\bar u 
\int_0^1 d\lambda
G_0(0,1;\bar u,\lambda) \frac{\sigma_D^2(\bar u\sigma_0\sqrt{T}) - \sigma_0^2}{\sigma_0^2} \\
& &\times
\Big(-\frac{1}{2\lambda} + \frac{(\bar u-\bar y)^2}{2\lambda^2}\Big) G_0(\bar u,\lambda; \bar y,0) 
(\bar y)_+ \nonumber
\end{eqnarray}
A similar rescaling gives for the zeroth order term
\begin{eqnarray}
C_0(0,T) = \sigma_0 \sqrt{T}
\int_{-\infty}^\infty d\bar y
G_0(\bar u,\lambda; \bar y,0) (\bar y)_+
\end{eqnarray}
We observe that the overall scaling factor $\sqrt{T}$ is the same for $C_0$ and $C_1$, and the only  
new time dependence in $C_1$ appears from the
factor involving the local volatility $\sigma_D^2(\bar u\sigma_0\sqrt{T})$. Assuming analyticity,
Taylor expansion of this factor produces both integer and half-integer powers of time
\begin{eqnarray}\label{sigDTexp}
\sigma_D^2(\bar u\sigma_0\sqrt{T}) - \sigma_0^2 = \sigma_0\sqrt{T} \bar u (\sigma_D^2(0))' + 
\frac12 (\sigma_0\sqrt{T})^2 \bar u^2 (\sigma_D^2(0))'' + \cdots 
\end{eqnarray}

\begin{table}
\caption{\label{PTtable} 
Contributions to the ATM implied volatility from the $n$-th order of perturbation in $\Delta\hat L$.}
\begin{center}
\begin{tabular}{|c|r|}
\hline 
$n$ &  ATM implied vol \\
\hline \hline
1 &  $T^\frac12, T, T^\frac32, T^2, \cdots$ \\
2 &  $T, T^\frac32, T^2, \cdots$ \\
3 &  $T^\frac32, T^2, \cdots$ \\
4 &  $T^2, \cdots$ \\
\hline \hline
\end{tabular}
\end{center}
\end{table}

At this point we stop and examine the structure of the time-dependent terms at higher orders in 
perturbation theory. The $n$-th order in perturbation theory $C_n(K,T)$ will be given by a 
$2n+1$-dimensional
integral, which includes $n$ double integrals over intermediate space-time variables, plus 
one final integral over the initial condition. The time dependence will consist of an overall 
$\sigma_0\sqrt{T}$ factor as in $C_{0,1}(K,T)$ plus terms arising from the presence of $n$ factors 
of the form (\ref{sigDTexp}). The resulting contributions at each order $n$ in perturbation theory 
can be easily obtained and are shown in Table \ref{PTtable}.

It is clear that only the first order in perturbation theory can contribute a term of $O(\sqrt{T})$ 
to the ATM implied volatility. However, upon computing the integrals over $\bar u, \bar y$ we get
a vanishing result
\begin{eqnarray}
&& \int_0^\infty d\bar y \bar y \int_{-\infty}^\infty d\bar u 
G_0(0,1;\bar u,\lambda) \bar u (-\frac{1}{2\lambda} + \frac{(\bar u -\bar y)^2}{2\lambda^2}) G_0(\bar u,\lambda;\bar y,0) = 
 \int_0^\infty d\bar y \bar y^2 e^{-\frac12 \bar y^2} (3-\bar y^2) = 0 \,.\nonumber \\
\end{eqnarray}
This proves the absence of a $O(\sqrt{T})$ term in the ATM implied volatility for the case
of an analytical local volatility. This is in agreement with the results of Sections 2, 3, 
which considered only this case. 

However, the situation is very different if we assume a nonanalytical local volatility.
Taking an expansion for $\sigma_D^2(\bar u \sigma_0 \sqrt{T})$ with a linear term of the form
\begin{eqnarray}
\sigma_D^2(\bar u \sigma_0 \sqrt{T}) - \sigma_0^2 = 
\left\{
\begin{array}{cc}
\bar u \sigma_0 \sqrt{T} b_R & \mbox{if } \bar u > 0 \\
\bar u \sigma_0 \sqrt{T} b_L & \mbox{if } \bar u \leq 0 \\
\end{array}
\right.
\end{eqnarray}
the integration over $\bar u, \bar y$ does not vanish anymore, and the result is proportional
with the discontinuity of the derivative $b_R - b_L$
\begin{eqnarray}
C_1(0,T) = \sigma_0 \sqrt{T} \Big( \frac{1}{32}\sigma_0 \sqrt{T} (b_R - b_L) \Big)
\end{eqnarray}
Using the relation $C(0,T) = \frac{1}{\sqrt{2\pi}} \sigma_N(0,T)\sqrt{T}$, we obtain from this
result the $O(\sqrt{T})$ term in the ATM implied volatility.
This concludes the proof of the relation (\ref{sqrtTterm}).

The explicit result for the normal implied volatility Eq.~(\ref{sig22B}) for the 
model (\ref{2B}) contains only one non-analytic term $\sim \sqrt{T}$, but not other
terms of similar form, e.g. $T^{3/2}, T^{5/2}, \cdots$. It would be interesting
to investigate whether this is a general result, or if it holds only in the specific
model (\ref{2B}). 

Although the relation (\ref{sqrtTterm}) was proved for the ATM normal volatility, a similar result
holds also for the ATM log-normal volatility. This is related to the former by the
exact relation (\ref{ATMN2LN})
\begin{eqnarray}
\sigma_N(0,T) = F_T
\sqrt{\frac{2\pi}{T}} \mbox{Erf}\Big(\frac{\sigma_{BS}(0,T)\sqrt{T}}{2\sqrt2}\Big)
\end{eqnarray}
In the short-maturity limit $T\to 0$ this reduces to the well-known relation 
$\sigma_N(0) = F_T \sigma_{BS}(0)$. The relation between $\sigma_N(0,T)$ and
$\sigma_{BS}(0,T)$ is an analytical function of time, which implies that a
term proportional to $\sqrt{T}$ in the expansion of the ATM normal implied 
volatility will introduce such a contribution also in the ATM log-normal implied
volatility.

\begin{figure}
    \centering
   \includegraphics[width=3in]{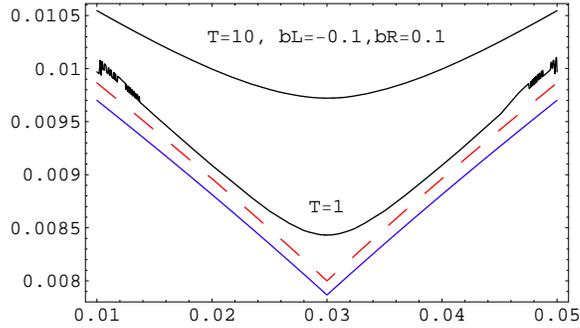}
    \caption{
The normal implied volatility $\sigma_N(K)$ for the model (\ref{2B})
comparing the exact solution (black) with the asymptotic solution $\sigma_0(K)$
(red dashed), and the solution including $O(T)$ corrections (solid blue).
Model parameters: $S=3\%, \sigma_0 = 0.8\%, T=10$. $b_{L,R}$ are as shown in the plots. }
 \end{figure}

\begin{figure}
    \centering
   \includegraphics[width=3in]{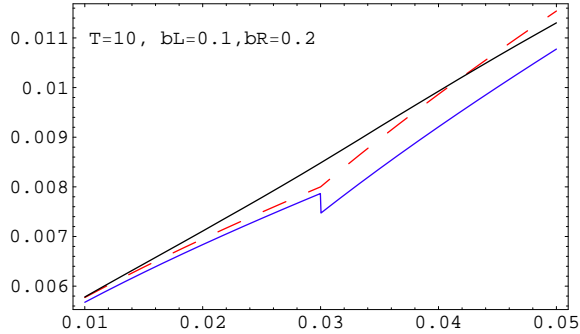}
    \caption{
Discontinuity of the asymptotic solution at $O(T)$ for the model (\ref{2B}),
with $|b_L| \neq |b_R|$. As explained in text, the subleading correction $\sigma_1(y)$ is 
discontinuous at $y=0$.
The black curve shows the exact (numerical) solution, the red dashed curve shows the
asymptotic solution $\sigma_0(K)$, and the blue curve shows the asymptotic
solution including $O(T)$ corrections.
Model parameters: $S=3\%, \sigma_0 = 0.8\%, T=10, b_L=0.1, b_R=0.2$. }
 \end{figure}

Another manifestation of the failure of the asymptotic expansion for non-analytic
local volatility functions is the prediction of a discontinuity in the implied volatility.
This can happen for example in the model (\ref{2B}) at order $O(T)$ in the
$T$ expansion, provided that $|b_L| \neq |b_R|$. The correction of $O(T)$ to
the normal implied volatility is given by Eq.~(\ref{sig1K}).
The jump of the 
subleading correction can be obtained from the power series expansion of
$\sigma_1(y)$ in the shifted log-normal model, and is 
\begin{eqnarray}
\sigma_1(0+\epsilon) - \sigma_1(0-\epsilon) = -\frac16 (b_R^2 - b_L^2)\sigma_0
\end{eqnarray}
This discontinuity is shown in Figure~3 for a specific choice of the
model parameters. Such a discontinuity is clearly an artifact of the
naive application of the asymptotic expansion, as the numerical solution is 
continuous everywhere.

\section{Conclusions}

In this paper we extended the short time asymptotic expansion method to 
local volatility models described in terms of normal volatility, as opposed to
log-normal volatility. A description in terms of
normal implied volatility appears naturally in the context of interest rates, 
which can become negative in regimes of small interest rates. The Dupire equation
can be formulated as a nonlinear equation for the implied normal volatility,
which can be solved by an asymptotic expansion in powers of time $T$.
This equation is similar (although in a different independent variable - the 
difference strike $y=K-F_T$ as opposed to the log-strike $x=\log(K/F_T)$)
to the usual BBF equation \cite{BBF}, from which it differs only at $O(T^2)$.

We present explicit solutions for the coefficients of the $O(T), O(T^2)$ 
terms in the small-time expansion of the normal implied volatility. The
drift term in the $O(T)$ coefficient can be expressed as a simple integral
over the zeroth order coefficient and the local volatility. We point out that
the drift contribution vanishes at the ATM point $K=S_0$ in the $O(T)$ term.
This absence is verified on the explicit example of the shifted log-normal model
with constant drift, for which an exact solution can be obtained.

We studied the convergence of the asymptotic expansion on two examples of
analytical local volatility: shifted log-normal model with and without drift,
and a model inspired from stochastic volatility models for which the local 
volatility is the square root of a quadratic polynomial of strike. We found 
generally good agreement, even when keeping only the first two terms $O(1), O(T)$
in the small-time expansion. 

The asymptotic expansion of the implied volatility in integer powers of $T$ 
can fail if the local volatility is a nonanalytic function. Using perturbation 
theory techniques, we show that
in local volatility models where the local volatility has a discontinuous
derivative at the ATM point, the ATM normal implied volatility contains 
nonanalytic dependence on time, proportional to $\sqrt{T}$. Furthermore,
the coefficient of this term is simply determined, and is proportional to the
jump of the derivative at the nonanalyticity point. Although these results have
been proven for the normal implied volatility, similar results must hold also 
for the log-normal implied volatility, which is related to the former by an
analytical function of time. 

Another manifestation of the failure of the usual
asymptotic expansion for such models is the presence of a jump discontinuity
in the $O(T)$ contribution to the implied volatility. These observations show 
that the usual asymptotic expansion can fail for non-analytic local volatility 
functions, and has to be used with care in such situations.

\newpage
\appendix

\section{Relation between the normal and log-normal implied volatilities}

We quote here an exact relation between normal and log-normal implied volatilities valid for
arbitrary maturity $T$, but for small deviations from the ATM point $y = K-F_T \ll F_T$.
This takes the form of an expansion in $y/F_T$, and the first few terms are
\begin{eqnarray}
w_N(y,T) &=& F_T \left\{
\sqrt{2\pi} \mbox{Erf}\Big(\frac{w}{2\sqrt2}\Big) +
\sqrt{\frac{\pi}{2}}  \mbox{Erf}\Big(\frac{w}{2\sqrt2}\Big) \frac{y}{F_T}\right. \\
 &+& \left.
\Big(  \frac{1}{2w} e^{-w^2/8} - \frac{1}{2\sqrt{2\pi}\mbox{Erf}\Big(\frac{w}{2\sqrt2}\Big)}
\Big) \Big(\frac{y}{F_T}\Big)^2 + O\Big(\frac{y}{F_T}\Big)^3 \right\}\nonumber
\end{eqnarray}
We denoted here $w_N(y,T) = \sigma_N \sqrt{T}$ and $w(y,T) = \sigma_{BS} \sqrt{T}$ the 
square roots of the normal and log-normal variances, respectively. 

The ATM implied volatilities satisfy the relation
\begin{eqnarray}\label{ATMN2LN}
\sigma_N(0,T) = F_T
\sqrt{\frac{2\pi}{T}} \mbox{Erf}\Big(\frac{\sigma_{BS}(0,T)\sqrt{T}}{2\sqrt2}\Big)
\end{eqnarray}
which reduces in the short-maturity limit $T\to 0$ to the well-known relation 
$\sigma_N(0) = F_T \sigma_{BS}(0)$.

Taking the derivative with respect to strike at the ATM point gives an exact relation
between the normal and log-normal skews
\begin{eqnarray}
\sigma'_N(0,T) = F_T [ \sigma'_{BS}(0,T) + \sqrt{\frac{\pi}{2}} \frac{1}{\sqrt{T}}
\mbox{Erf}\Big( \frac{\sigma_{BS}(0,T)\sqrt{T}}{2\sqrt2}\Big)]
\end{eqnarray}
which reduces in the short time limit to 
\begin{eqnarray}\label{ATMexpand}
\sigma'_N(F_T) &=& \frac12 \sigma_{BS}(F_T) + F_T \sigma'_{BS}(F_T)  
\end{eqnarray}

The asymptotic results of Sec.~2 give an exact 
relation between the normal and log-normal implied volatilities, valid in the short-time 
asymptotic limit. We limit ourselves to the case of time-homogeneous local volatility. 
The relations (\ref{BBF}) and (\ref{BBFN}) express both of them in terms of the same 
local volatility $\sigma_D(x)$. Eliminating the local volatility gives a direct relation 
between the two types of implied volatilities. 

The starting point is Eq.~(\ref{BBFN}), where we substitute the local volatility $\sigma_D(y)$ with its expression in terms of the log-normal
volatility
\begin{eqnarray}
\sigma_D(x) = \frac{\sigma_{BS}(x)}{1 - \frac{x}{\sigma_{BS}(x)}\partial_x\sigma_{BS}(x)}
\end{eqnarray}
with $x = \log(K/F_T)$.

The integral in the denominator of (\ref{BBFN}) can be performed by integration by parts
\begin{eqnarray}
\int_{F_T}^K dL \frac{1}{\sigma_D(L)} = \int_0^x \frac{dy}{\sigma_{BS}(y)} - \int_0^x \frac{y}{\sigma^2_{BS}(y)}d\sigma_{BS} = \frac{x}{\sigma_{BS}(x)}
\end{eqnarray}
and we find the very simple result
\begin{eqnarray}
\sigma_N(K) = \sigma_{BS}(K) \frac{K - F_T}{\log\frac{K}{F_T}}
\end{eqnarray}

Expanding around this point one can find relations among the
skews and curvatures (convexities) of the normal and log-normal ATM skews, valid in the
short time limit. We quote here the relation among convexities
\begin{eqnarray}
\sigma''_N(F_T) &=& -\frac{1}{6F_T}\sigma_{BS}(F_T) +   \sigma'_{BS}(F_T) +  F_T \sigma''_{BS}(F_T)
\end{eqnarray}

\section{Log-normal process with constant drift}
\label{sec:appB}

As an explicit example of local volatility model of the type (\ref{sigDN}) with 
non-zero drift, we consider the simplest such process, with log-normal volatility
and constant drift
\begin{eqnarray}\label{lnmu}
dx_t = x_t dW_t + \mu dt
\end{eqnarray}
We will compute the normal implied volatility in this model, and compare its small time
expansion
with the asymptotic expansion derived in Sec.~\ref{sec:3}. As noted in Sec.~\ref{sec:4},
this explicit example confirms the correctness of 
the $O(T)$ correction to the normal implied volatility (\ref{sig1K}).

The stochastic differential equation (\ref{lnmu}) can be integrated in closed form 
with the result
\begin{eqnarray}
x_t = x_0 \exp(W_t - \frac12 t) + \mu \int_0^t ds \exp(W_t-W_s-\frac12(t-s))\,.
\end{eqnarray}
From this expression it is clear that $x_t$ is restricted to the range
$x_t > 0$, provided that $\mu\geq 0$. In the following we will restrict ourselves 
to the case $\mu \geq 0$.

The mean and variance of $x_t$ are
\begin{eqnarray}\label{meanvarx}
&& E[x_t] = x_0 + \mu t \,,\\
&& \mbox{var}(x_t) = E[x_t^2] - E[x_t]^2 = x_0^2 (e^t-1)+
2x_0 \mu (e^t - 1 - t) + 
2\mu^2 (e^t - 1 - t - \frac{t^2}{2})\,.\nonumber
\end{eqnarray}

The process for $z_t = \log(x_t)$ is mean-reverting around 
$z_{\rm mid} = \log(2\mu)$.
This reads $dz_t = dW_t + (\mu e^{-z_t} - 
\frac12) dt$, which is similar to the Ornstein-Uhlenbeck process, 
but with an exponential mean-reverting term. From Eq.~(\ref{meanvarx}) we see
that the variance of $x_t$ increases with $t$ and is unbounded, which is 
different from the OU proces, for which it approaches a finite limit as $t\to \infty$.

We would like to compute the price of a call option on $x_t$ with strike $K$ 
conditional on $x_0$ at $t=0$
\begin{eqnarray}
C(K,t) = E[(x_t-K)_+|x_0,t=0]
\end{eqnarray}
It satisfies the Dupire equation
\begin{eqnarray}\label{Deq}
\partial_t C(K,t) = \frac12 K^2 \partial_K^2 C(K,t) - \mu \partial_K C(K,t)
\end{eqnarray}
with the boundary conditions
\begin{eqnarray}\label{BC}
&& C(K,0) = (x_0 - K)_+ \\
&& C(0,t) = x_0 + \mu t\,,\qquad \partial_K C(K,t)|_{K=0} = -1\,,\qquad
\lim_{K\to \infty} \partial_K C(K,t) = 0\,.
\end{eqnarray}

We will solve the Dupire equation (\ref{Deq}) using Laplace transform methods.
Define the Laplace transform of the call price with respect to time
\begin{eqnarray}
y(x,s) = \int_0^\infty dt e^{-st} C(x,t)
\end{eqnarray}
This satisfies the Dupire equation in Laplace transformed form
\begin{eqnarray}\label{DL}
\frac12 x^2 y''(x) - \mu  y'(x) - s y(x) &=& x - x_0\,,\qquad 0 \leq x \leq x_0 \\
&=& 0 \,,\qquad\qquad\quad x > x_0 \nonumber 
\end{eqnarray}
with boundary conditions
\begin{eqnarray}\label{BCL}
y(0,s) = \frac{x_0}{s} + \frac{\mu}{s^2}\,,\qquad
y'(0,s) = -\frac{1}{s}\,,\qquad
\lim_{x\to \infty} y'(x,s) = 0\,.
\end{eqnarray}

The solution of (\ref{DL}) is given by the sum of a particular solution of the 
inhomogeneous equation plus the most general solution of the homogeneous equation
\begin{eqnarray}\label{133}
y(x) &=&  \left\{
\begin{array}{cc}
 - \frac{1}{s}(x - x_0 - \frac{\mu}{s}) + c_1 f_1(x) + c_2 f_2(x)  & x < x_0 \\
  d_2 f_2(x) & x > x_0 \\
\end{array}
\right. 
\end{eqnarray}
where $f_{1,2}(x)$ are the solutions of the homogeneous equation. They are
\begin{eqnarray}
f_1(x) &=&  x^{\lambda_1} {}_1 F_1(-\lambda_1 , 2\lambda_2; - \frac{2\mu}{x}) \\
f_2(x) &=&  x^{\lambda_2} {}_1 F_1(-\lambda_2 , 2\lambda_1; - \frac{2\mu}{x}) 
\end{eqnarray}
with ${}_1 F_1(a,b,z)$ the confluent hypergeometric function, and
$\lambda_{1,2}$ are functions of $s$
\begin{eqnarray}
\lambda_1 = \frac12 (1 + \sqrt{1 + 8s}) > 0\,,\qquad 
\lambda_2 = \frac12 (1 - \sqrt{1 + 8s}) < 0
\end{eqnarray}

In the $x>x_0$ region we kept only the second solution $f_2(x)$ since it is
the only one which satisfies the condition $y'(x) \to 0$ as $x\to \infty$.
For large values of the argument $x$ the asymptotics of the solutions is
\begin{eqnarray}
f_1(x) \simeq x^{\lambda_1}\,,\qquad 
f_2(x) \simeq x^{\lambda_2}
\end{eqnarray}
such that $f_1(x)$ is increasing while $f_2(x)$ decreases approaching 0 as
$x\to \infty$.

The wronskian of the two solutions of the homogeneous equation $f_{1,2}(x)$ is
\begin{eqnarray}\label{wronskian}
f_1(x) f'_2(x) - f'_1(x) f_2(x) = (\lambda_2 - \lambda_1)\exp( -\frac{2\mu}{x})=
- \sqrt{1+8s}\exp( -\frac{2\mu}{x})\,.
\end{eqnarray}
The boundary conditions at $x=0$ (\ref{BCL}) require that the constants $c_{1,2}$
satisfy the conditions
\begin{eqnarray}\label{cond1}
&& c_1 f_1(0) + c_2 f_2(0) = 0 \\
&& c_1 f'_1(0) + c_2 f'_2(0) = 0 \,.
\end{eqnarray}
This will have nonzero solutions for $c_{1,2}$ only if the wronskian of the
two functions vanishes at $x=0$. Using the explicit expression in (\ref{wronskian}) 
this is seen to be indeed the case. Only one of the two equations above is independent;
we will choose the first one.

The limit of the functions $f_{1,2}(x)$ at $x=0$ is
\begin{eqnarray}
f_1(0) = (2\mu)^{\lambda_1} \frac{\Gamma(2\lambda_2)}{\lambda_2 \Gamma(\lambda_2)}\,,\qquad
f_2(0) = (2\mu)^{\lambda_2} \frac{\Gamma(2\lambda_1)}{\lambda_1 \Gamma(\lambda_1)}
\end{eqnarray}

Another condition on the constants $c_{1,2}, d_2$ follows from
requiring continuity of $y(x)$ at $x=x_0$ 
\begin{eqnarray}\label{cont}
&& c_1 f_1(x_0) + c_2 f_2(x_0) - d_2 f_2(x_0) = -\frac{\mu}{s^2}
\end{eqnarray}

Finally, another condition is obtained from the normalization of the integrated
pdf
\begin{eqnarray}
\int_0^\infty dx y''(x) = \int_0^{x_0} dx (c_1 f''_1(x) + c_2 f''_2(x)) +
\int_{x_0}^\infty dx d_2 f''_2(x) = \frac{1}{s}
\end{eqnarray}
which gives
\begin{eqnarray}
&& c_1 f'_1(x_0) + c_2 f'_2(x_0) - d_2 f'_2(x_0) = \frac{1}{s}
\end{eqnarray}

Together with (\ref{cond1}) and (\ref{cont}) this equation fixes the constants
$c_{1,2}, d_2$. These constants are found by solving the equations
\begin{eqnarray}
&& c_1 f_1(0) + c_2 f_2(0) = 0 \\
&& c_1 f_1(x_0) + c_2 f_2(x_0) - d_2 f_2(x_0) = -\frac{\mu}{s^2} \\
&& c_1 f'_1(x_0) + c_2 f'_2(x_0) - d_2 f'_2(x_0) = \frac{1}{s}
\end{eqnarray}
The solution is
\begin{eqnarray}\label{c1sol}
c_1 &=& \frac{f_2(x_0) + (\mu/s) f'_2(x_0)}{(f'_1(x_0) f_2(x_0) - f_1(x_0) f'_2(x_0)) s} =
\frac{f_2(x_0) + (\mu/s) f'_2(x_0)}{s\sqrt{1+8s}}\exp(\frac{2\mu}{x_0})\\
\label{c2sol}
c_2 &=& -\frac{f_1(0) [f_2(x_0) + (\mu/s) f'_2(x_0)]}{f_2(0)(f'_1(x_0) f_2(x_0) - f_1(x_0) f'_2(x_0)) s}\\
&=&
(2\mu)^{\sqrt{1+8s}}\frac{\lambda_1\Gamma(\lambda_1)\Gamma(2\lambda_2)}{\lambda_2\Gamma(\lambda_2)\Gamma(2\lambda_1)}
\frac{[f_2(x_0) + (\mu/s) f'_2(x_0)]}{s\sqrt{1+8s} }\exp(\frac{2\mu}{x_0})\nonumber\\
\label{d2sol}
d_2 &=& \frac{f_1(0) f_2(x_0) - f_1(x_0) f_2(0) + [f_1(0) f'_2(x_0) - f'_1(x_0) f_2(0)] (\mu/s)}
{f_2(0) (f_1(x_0) f'_2(x_0) - f'_1(x_0) f_2(x_0)) s}\,.
\end{eqnarray}
This completes the solution of the Laplace transformed Dupire equation (\ref{DL}).
The call price $C(K,t)$ is obtained by taking the inverse Laplace transform of
the solution $y(x,s)$ given in Eq.~(\ref{133}) with the coefficients $c_{1,2}, d_2$ given
in (\ref{c1sol}), (\ref{c2sol}), (\ref{d2sol}).

We consider next the explicit result for $y(x_0,s)$, which is the Laplace 
transform over time of the ``ATM'' call price (the quotation marks are a reminder 
that this is the ATM point only for the case of zero drift; in the presence
of the drift the ATM point corresponds to $x=x_0 + \mu/s$)
\begin{eqnarray}
y(x_0,s) &=& \frac{\mu}{s^2} + \frac{\exp(2\mu/x_0)}{s\sqrt{1+8s}} [f_2(x_0)+\frac{\mu}{s} f'_2(x_0)]
[f_1(x_0) - \frac{f_1(0)}{f_2(0)} f_2(x_0)] \\
&=& \frac{\mu}{s^2}\Big(1 + \frac{\exp(2\mu/x_0)}{\sqrt{1+8s}}\Big) f_1(x_0) f'_2(x_0)) \nonumber\\
& & + 
x_0 \frac{\exp(2\mu/x_0)}{s\sqrt{1+8s}} 
{}_1 F_1(-\lambda_1 , 2\lambda_2; - \frac{2\mu}{x_0})\, {}_1 F_1(-\lambda_2 , 2\lambda_1; - \frac{2\mu}{x_0})
+ \cdots
\end{eqnarray}
where the ellipses are proportional to factors of $(2\mu)^{\sqrt{1+8s}}$ and $(2\mu)^{1+\sqrt{1+8s}}$.

This is easily expanded in powers of the drift $\mu$. Keeping only terms linear in $\mu$ we find
\begin{eqnarray}
y(x_0,s) &=& x_0 \frac{1}{s\sqrt{1+8s}}
\Big( 1 + \frac{\mu}{x_0} \frac{(\lambda_1 + \lambda_2)^2}{\lambda_1\lambda_2} + O(\mu^2) \Big) + 
\frac{\mu}{s^2}(1 + \frac{\lambda_2}{\sqrt{1+8s}})\\
&=& x_0 \frac{1}{s\sqrt{1+8s}}\Big( 1 - \frac{\mu}{x_0} \frac{1}{2s} + O(\mu^2) \Big) + 
\frac{\mu}{s^2}(\frac12 + \frac{1}{2\sqrt{1+8s}})
\end{eqnarray}

The first term corresponds to the usual Black-Scholes result for the ATM call. Its inverse Laplace transform can be found easily and is
\begin{eqnarray}
\frac{1}{s\sqrt{1+8s}} \to \mbox{Erf} \Big(\frac{\sqrt{t}}{2\sqrt2}\Big)
\end{eqnarray}
Its expansion in powers of $t$ contains $\sqrt{t}, 1/\sqrt{t}, \cdots$.


There is a partial cancellation of the  $O(\mu)$ terms, such that we get the ``ATM'' call price 
(in Laplace transformed form)
\begin{eqnarray}
y(x_0,s) = x_0 \frac{1}{s\sqrt{1+8s}} + \frac{\mu}{2s^2} + O(\mu^2, \mu^{\sqrt{1+8s}})
\end{eqnarray}
The inverse Laplace transform of this result is easily computed, and we find
the explicit expansion of the ``ATM'' call price in powers of the drift
\begin{eqnarray}\label{final}
C(K=x_0,t) &=& x_0 \mbox{Erf} (\frac{\sqrt{t}}{2\sqrt2}) + \frac12 \mu t + O(\mu^2)\\
 &=&
\frac{\sqrt{t}}{\sqrt{2\pi}} 
x_0 \Big(1- \frac{t}{24} + \frac{t^2}{640} - \frac{t^3}{21504} + \cdots \Big) + \frac12 \mu t
+ O(\mu^2)\,.\nonumber
\end{eqnarray}

We verified that the same result is obtained also using perturbation methods for linear
operators applied to the Dupire equation (\ref{Deq}), considering the drift term as a 
perturbation.

\section{Proof of the relation (\ref{ATM2B})}

We present here the details for the exact solution of the local volatility
model with the local volatility
\begin{eqnarray}\label{2Beq}
\sigma_D(y) = \left\{
\begin{array}{cc}
\sigma_0 - 2b y\,, & y < 0 \\
\sigma_0 + 2b y\,, & y > 0 \\
\end{array}\right.
\end{eqnarray}

Define the new variable
\begin{eqnarray}
z(y) = \int_0^y \frac{du}{\sigma_D(u)} 
\end{eqnarray}
which follows the process
\begin{eqnarray}\label{zproc}
dz(t) = dW(t) - b(z) dt\,, \qquad
b(z) = \left\{
\begin{array}{cc}
-b\,, & z < 0 \\
b\,, & z > 0 \\
\end{array}\right.
\end{eqnarray}
The original variable $y$ is expressed
in terms of $z$ as
\begin{eqnarray}
y = \left\{
\begin{array}{cc}
-\frac{\sigma_0}{2b}( e^{-2bz} - 1)\,, & z < 0 \\
\frac{\sigma_0}{2b}( e^{2bz} - 1)\,, & z > 0 \\
\end{array}\right.
\end{eqnarray}

The fundamental solution of the process (\ref{zproc}) is known exactly, see
Eq.~(5.14) in \cite{KS}
\begin{eqnarray}
p(z,t;x,0) &=& \frac{1}{\sqrt{2\pi t}}\Big[
\exp(- \frac{(x-z-bt)^2}{2t}) + b e^{-2bz} 
\int_{x+z}^\infty dv \exp( - \frac{(v-bt)^2}{2t}) 
\Big]\,,\quad x\geq 0, z>0\nonumber \\
&=& \frac{1}{\sqrt{2\pi t}}\Big[
\exp(2bx - \frac{(x-z+bt)^2}{2t}) + b e^{2bz} 
\int_{x-z}^\infty dv \exp( - \frac{(v-bt)^2}{2t}) 
\Big]\,,\quad x\geq 0, z\leq 0 \nonumber\\
\end{eqnarray}

It is easy now to compute prices of call options using the relation
\begin{eqnarray}
C(S_0,K,t) = \int_{z(K)}^\infty du (y(u) - K) p(u,t;z(S_0),0)
\end{eqnarray}
In particular, the ATM implied normal volatility in Eq.~(\ref{ATM2B}) 
was obtained from the ATM call price according to
\begin{eqnarray}
C(S_0,S_0,t) = \frac{1}{\sqrt{2\pi}} \sigma_{ATM}(S_0) \sqrt{t}
\end{eqnarray}

\vspace{0.3cm}\noindent
{\em Acknowledgments:}
D.~P. thanks Jim Gatheral for stimulating comments and advice about the local volatility models, and 
Bruno Dupire for a discussion. V.~C. thanks Alan Lewis for correspondence
on the numerical solution of partial differential equations. We thank Radu Constantinescu for
discussions, advice and comments.

\end{document}